\documentclass[11pt,a4paper]{article}

\usepackage[a4paper,margin=2.5cm]{geometry}

\usepackage{amssymb,amsmath,amsthm}
\usepackage[T1]{fontenc}
\usepackage[utf8]{inputenc}
\usepackage{lmodern,microtype}
\usepackage{graphicx}
\usepackage{xcolor}
\usepackage{xspace}
\usepackage{bm}
\usepackage{url}
\usepackage{multirow}
\usepackage{booktabs}
\usepackage[export]{adjustbox}
\usepackage[hidelinks]{hyperref}
\usepackage[linesnumbered,ruled,vlined,noend]{algorithm2e}

\newtheorem{theorem}{Theorem}
\usepackage{enumitem}
\newtheorem{lemma}{Lemma}

\newcommand{\curN}{{n_{\mathrm{cur}}}\xspace}

\newcommand{\poly}[1]{{\texttt{poly}({#1})}}

\newcommand{\TNI}{m\xspace}
\newcommand{\estimateFunction}{\textnormal{\textsc{Estimate}}\xspace}

\newcommand{\LatRV}{\mathsf{L}}
\newcommand{\Service}{\mathrm{Service}}

\newcommand{\TSTB}{\textnormal{\textsc{Truncated Sawtooth-Backoff}}\xspace}

\newcommand{\tstb}{\textnormal{\textsc{TSTB}}\xspace}

\newcommand{\Cost}{\mathrm{Cost}\xspace}

\newcommand{\optimalAlgo}{\textsc{OPT}\xspace}

\newcommand{\mainAlgorithm}{\textsc{Randomized Adaptive Channel Exploitation}\xspace}
\newcommand{\mainAlgorithmAcrynom}{\textnormal{\textsc{RACE}}\xspace}

\newcommand{\defn}[1]{\textbf{\emph{#1}}}

\hypersetup{pdftitle={Contention Resolution with a Fast-but-Fickle Channel}}

\title{Contention Resolution with a Fast-but-Fickle Channel}
\author{
Umesh Biswas, Vini Chaudhary, Shafqat Hasan, Maxwell Young\\[6pt]
Mississippi State University\\
Mississippi State, Mississippi, USA\\[4pt]
\texttt{ucb5@msstate.edu}, \texttt{vchaudhary@cse.msstate.edu}\\
\texttt{sh2924@msstate.edu}, \texttt{myoung@cse.msstate.edu}
}
\date{}

\newtheorem{proposition}{Proposition}
\newcommand{\ceil}[1]{\left\lceil #1\right\rceil}

\begin{document}
\maketitle

\begin{abstract}
The contention-resolution problem studies how {\boldmath{$n$}} devices share a communication channel without centralized coordination. Classical models typically assume a single continuously available channel and focus on minimizing total completion time. However, modern wireless systems increasingly rely on shared spectrum, where access may be abruptly revoked by incumbent users.

Motivated by this setting, we study contention resolution with two channels: a reliable slow channel and a faster channel that can be intermittently disabled by unpredictable eviction periods. Moving a batch of packets from the slow channel to the fast channel incurs a unit cost, while evictions from the fast channel impose a larger per-packet penalty. Our goal is to achieve asymptotically optimal latency while minimizing switching overhead relative to an optimal offline algorithm that knows the eviction schedule.

We present \mainAlgorithm\ (\mainAlgorithmAcrynom), a randomized algorithm that exploits the fast channel when available while maintaining progress on the slow channel during disruptions. When the fast channel has rate $\rho \in \bigl[\Omega(\log n), O(\sqrt n)\bigr]$ and a sufficiently large constant-factor resource augmentation, \mainAlgorithmAcrynom\ achieves an $O(1)$ competitive ratio for expected latency. \mainAlgorithmAcrynom's expected switching-cost competitive ratio relative to any optimal-latency offline algorithm \optimalAlgo is $O(\rho+\sqrt{n/\TNI})$, where $\TNI\geq 1$ is the number of fast-channel availability intervals used by OPT.

We also prove a worst-case $\Omega(\sqrt n)$ switching-cost competitive ratio for latency-competitive randomized algorithms, and a separate $\Omega(\rho)$ switching-cost competitive-ratio lower bound. These lower bounds hold even with fixed constant-factor augmentation on both channels.

\end{abstract}

\noindent\textbf{Keywords:} Contention resolution, online algorithms, multiple access channel, resource augmentation, distributed algorithms.

\medskip

\section{Introduction}\label{sec:introduction}

Dynamic spectrum sharing offers a way to increase wireless capacity. Rather than assigning spectrum exclusively, underutilized government-licensed spectrum can be shared with commercial and industrial users \cite{bhattarai2016overview,peha1998spectrum,calabrese2021useitorshare}. These secondary users may access spectrum opportunistically, but must avoid interference with incumbents, who have priority access rights. For example, the U.S. Citizens Broadband Radio Service (CBRS) framework allows secondary access while protecting incumbents \cite{fcc_sas_35ghz}. Current Spectrum Access System (SAS) deployments are centralized, though decentralized architectures have been proposed to improve scalability and robustness \cite{xiao2022decentralized}.

A central algorithmic difficulty is that secondary access can be revoked. When an incumbent becomes active, secondary users may be forced to vacate the high-capacity spectrum; that is, they are evicted from the fast channel. Such evictions can be difficult to detect, predict, or coordinate around \cite{ida2015spectrum_sharing,caromi2019detection}. Thus, dynamic spectrum sharing provides high throughput but unpredictable availability.

Here, we address contention resolution in this setting. Contention resolution asks how packets, or the devices sending them, share a communication channel without centralized scheduling (see surveys \cite{banicescu2024survey,chlebus2001randomized}). A packet succeeds only when it transmits alone in a slot; if multiple packets transmit, a collision occurs and none succeeds. Classical models typically assume one continuously available channel, and the primary goal is to minimize latency, the time until all packets have successfully transmitted.

Dynamic spectrum sharing changes this picture. A secondary system may have access to a ``fast-but-fickle'' channel with high capacity, together with a slower fallback channel that remains continuously available. Algorithms must decide not only when packets transmit but also when they move between channels. These transitions incur overhead from re-association, reconfiguration, and service disruption. Excessive switching is therefore costly even when latency is low. The challenge is to exploit the fast channel when available, tolerate evictions, and avoid unnecessary switching.

Our work extends classical contention resolution to a two-channel model with a reliable slow channel and a faster channel subject to unpredictable evictions. We ask whether decentralized algorithms can achieve near-optimal latency while controlling the switching costs of moving packets between channels.\medskip

\medskip

\noindent{\bf Our Problem Statement.} Initially, $n$ packets share a reliable slow channel, each requiring one collision-free transmission to succeed. Packets may also use a faster channel subject to unpredictable evictions. We compare against an optimal offline algorithm, $\optimalAlgo$, that knows the eviction schedule and can schedule packet transmissions without collisions to minimize latency. Our goal is to design an online, decentralized algorithm whose expected latency is within a constant factor of $\optimalAlgo$'s while achieving a low switching-cost competitive ratio relative to $\optimalAlgo$.

\subsection{Model}\label{sec:our-model}

There are {\boldmath{$n$}} $\geq 2$ devices, each initially holding one packet. The \defn{population} is the number of packets not yet successfully transmitted; initially, it is $n$. Packets are anonymous and must transmit without centralized coordination. There are no later arrivals, and the algorithm does not initially know $n$. We describe actions as taken by packets rather than devices.\medskip

\noindent{\bf Communication Channels.}
Packets communicate over two multiple-access channels: the \defn{slow channel} and the \defn{fast channel}. Time is divided into \defn{slots}. On the slow channel, each packet may perform at most one action---either transmit or listen---per slot. On the fast channel, each slot consists of {\boldmath{$\rho$}} mini-slots, where $\rho\geq2$ is an integer, and a packet attached to the fast channel may transmit or listen in each mini-slot. Unless otherwise specified, \underline{``slot'' refers to a slow-channel slot}.

Each packet is attached to exactly one channel at a time, and all packets are initially attached to the slow channel. A packet may voluntarily switch from the slow channel to the fast channel only at a slow-slot boundary. In contrast, a packet may switch from the fast channel to the slow channel at any time, either voluntarily or as a result of an eviction. In either case, the packet becomes attached to the slow channel immediately but cannot act there until the next permitted slow-slot boundary.

In a slot, or in a fast mini-slot, no transmission produces an \defn{empty} outcome; exactly one produces a \defn{success}; and two or more produce a \defn{collision}, in which no transmission succeeds. Every packet participating on the relevant channel receives \defn{ternary feedback} indicating which of these three possible outcomes actually occurred. A transmitting packet immediately learns whether it succeeded and immediately terminates. A packet is \defn{active} until it succeeds, and $\curN$ denotes the current active population. Measurement and reporting transmissions also carry data: their successful senders terminate.\medskip

\noindent{\bf Evictions.}
As the online component of the model, the fast channel is subject to \defn{eviction periods}, during which packets cannot access it. An eviction period may begin at time zero and may last indefinitely. All packets on the fast channel when such a period begins are \defn{evicted}. At slot boundaries, packets on the slow channel can observe whether the fast channel is currently available. This observation requires no slow-channel action and reveals only whether the fast channel is available at that boundary.

The eviction schedule is fixed independently of the algorithm's random choices. Eviction periods may begin and end at arbitrary fast mini-slot boundaries; if an eviction begins exactly when a transmission ends, that transmission is completed before the eviction takes effect. An \defn{iteration} is a maximal interval of fast-channel availability. Every iteration that occurs contains at least $C_{\mathrm{avail}}$ full slow slots, for a sufficiently large fixed constant specified in Section~\ref{sec:execution-details}. We impose this restriction because very brief periods of fast-channel availability would offer little opportunity for meaningful performance gains, and rapid alternation between availability and eviction seems unlikely in practice. However, an eviction period may last indefinitely, so the fast channel is not guaranteed to become available again.\medskip

\noindent{\bf Metrics.}
We consider two cost measures:
\begin{itemize}[leftmargin=1.5em,itemsep=2pt,topsep=3pt]
\item \defn{Latency}: The number of slots until all packets succeed, rounded up to a slow-slot boundary.
\item \defn{Switching cost}: A batch of packets attempting to move from slow to fast incurs unit cost, even if the fast channel is unavailable. Moving $k$ packets from fast to slow costs $k$, whether voluntarily or by eviction.
\end{itemize}
Latency is the traditional performance measure for contention resolution \cite{banicescu2024survey,chlebus2001randomized}. Switching cost captures a separate concern: the overhead of changing channel assignments. Motivated by coordinated or group association in wireless systems \cite{omitola2018group,CAO20181,9383780}, our asymmetric model charges unit cost to move a batch onto the fast channel. Returning packets to the slow channel incurs a per-packet cost, reflecting the greater disruption and reconfiguration overhead of reassignment; for voluntary returns, this cost also captures the foregone opportunity to complete those packets' transmissions on the fast channel. 

We compare our algorithm to an optimal offline algorithm, \optimalAlgo, that knows the eviction schedule and schedules transmissions perfectly, without empty slots or collisions when usable service and pending packets are available. Let $\LatRV(\mathcal A,I)$ and $\Cost(\mathcal A,I)$ denote latency and switching cost on instance $I$; expectations are written explicitly for randomized algorithms. The minimum achievable latency under the same attachment and boundary rules is $\LatRV(\optimalAlgo,I)$. Fix a schedule attaining it, and let $\TNI$ be the number of iterations in which that schedule successfully transmits on the fast channel. Our upper bound holds for every such choice; the multiplicative switching comparison is asserted for $\TNI\geq1$. For the lower-bound denominator, we may choose a minimum-switching schedule among the minimum-latency schedules.

In the spirit of \defn{resource augmentation} \cite{roughgarden2020resource}, we assume our algorithm  has  $\lambda=O(1)$ \defn{logical subchannels} on the physical fast channel, for a sufficiently large fixed $\lambda$. Each is a noninterfering transmission resource providing one opportunity per mini-slot, with its own collisions and feedback. Thus our algorithm has aggregate capacity $\lambda\rho$ opportunities per slow slot, compared with $\rho$ for \optimalAlgo. An eviction affects all subchannels. \emph{There is no slow-channel augmentation: both algorithms have one slow opportunity per slot.} 

\subsection{Main Results}

Our results apply when $\rho=\Omega(\log n)$ and $\rho=O(\sqrt n)$, where the constant implicit in the lower bound is sufficiently large. Our decentralized algorithm \mainAlgorithm\ (\mainAlgorithmAcrynom) achieves the following guarantee.

\begin{theorem}\label{thm:main-theorem}
On every fixed instance $I$, \mainAlgorithmAcrynom guarantees
$$
\frac{E[\LatRV(\mainAlgorithmAcrynom,I)]}{\LatRV(\optimalAlgo,I)}=O(1).
$$
Moreover, if \optimalAlgo uses $\TNI\geq1$ fast-channel iterations, then
$$
\frac{E[\Cost(\mainAlgorithmAcrynom,I)]}{\Cost(\optimalAlgo,I)}
=O\left(\rho+\sqrt{\frac n\TNI}\right).
$$
\end{theorem}
The latency bound also applies when $\TNI=0$, and in this case, \mainAlgorithmAcrynom incurs zero switching cost with high probability and $O(1)$ expected switching cost.

The switching-cost bound improves when \optimalAlgo uses more fast-channel iterations. For example, when $\rho=\Theta(\log n)$ and $\TNI=\Theta(n/\poly{\log n})$, the ratio is $O(\poly{\log n})$.

Our upper bound is complemented by lower bounds:
\begin{theorem}\label{thm:lower-bounds}
Let $\mathcal A$ be any randomized online algorithm that is constant-competitive in expected latency. Then:
\begin{enumerate}[label=(\Alph*),itemsep=2pt,topsep=3pt]
\item There is a fixed eviction schedule on which $\mathcal A$ incurs expected switching-cost competitive ratio $\Omega(\sqrt n)$.
\item There is a fixed schedule in which a single eviction permanently disables the fast channel and $\mathcal A$ incurs expected switching-cost competitive ratio $\Omega(\rho)$.
\end{enumerate}
\end{theorem}
The first construction shows that our switching-cost competitive ratio is asymptotically tight when \optimalAlgo uses a constant number of fast-channel iterations: our upper bound becomes $O(\sqrt n)$, matching the $\Omega(\sqrt n)$ lower bound. The second shows that even a single eviction can force an $\Omega(\rho)$ switching-cost competitive ratio: exploiting the fast channel requires exposing sufficiently many packets to eviction, while OPT can avoid this cost by knowing the eviction time.

\section{Related Work}

We review related work on dynamic spectrum sharing, contention resolution, online scheduling, and resource augmentation.\smallskip

\noindent\textbf{Dynamic Spectrum Sharing.} Bhattarai et al. \cite{bhattarai2016overview} survey tiered-access architectures and database-driven coordination frameworks such as Spectrum Access Systems (SAS), which enable coexistence between incumbents and secondary users. AWS-3 and CBRS illustrate this approach \cite{ida2015spectrum_sharing}. Reliable incumbent detection using sensing and learning has also been studied \cite{caromi2019detection}, along with the regulatory and economic mechanisms underlying dynamic access \cite{peha1998spectrum,calabrese2021useitorshare}.

This literature largely emphasizes systems, experiments, regulation, and optimization. We complement it with a formal contention-resolution model and upper and lower bounds under unpredictable evictions.\medskip

\noindent\textbf{Static Contention Resolution.}
Classical contention resolution often studies the \defn{static} or \defn{batched-arrival} setting, in which all $n$ packets are initially present. The standard objective is makespan, the time until all packets succeed. Randomized and deterministic protocols have bounds ranging from $O(n\log n)$ for classical backoff approaches to optimal $O(n)$ makespan in suitable models \cite{bender2005adversarial,gereb1992efficient,greenberg1985randomized,komlos1985asymptotically,kowalski2005on}. Extensions consider adversarial arrivals \cite{borodin2001adversarial,chlebus2009maximum,chlebus2007stability,anantharamu2010deterministic,aldawsari2019adversarial}, energy efficiency \cite{jurdzinski2002energy,bender2016contention}, multiple slow channels \cite{christodoulou2018strategic}, and robustness to jamming \cite{bender2016how,awerbuch2008jamming,ogierman2014competitive,chen2021tight,anantharamu2019packet,anantharamu2011medium,bender2025jamming,bender2024fully}.

These models generally assume fixed communication resources, even under jamming. We instead study fast-channel availability interrupted by incumbent evictions, with switching cost as a second objective.

\smallskip

\noindent\textbf{Scheduling, Reallocation, and Shared Channels.}
Our model is also related to online scheduling, where packets correspond to jobs and channels to machines. A close classical analogue is scheduling on heterogeneous machines, including uniform or related machines with different speeds \cite{wen1998preemptive,liu_online_two_uniform,englert2021online}. Related directions include semi-online models with partial information \cite{dosa2011online,dosa2019tight,zheng2024semi} and job rejection, migration, or reassignment \cite{epstein2014preemptive,dosa2018tight,maack2023online}.

Several scheduling problems involve channels more explicitly: contention resolution with deadlines \cite{agrawal2020contention}, window scheduling and station assignment \cite{bar2003windows,bar2007windows,halper:station}, dynamic client reallocation across channels \cite{chan2004line,farach2021dynamic}, and fault-tolerant packet scheduling \cite{jurdzinski_packet_jamming,anta_speed_augmented_jamming,bohm2021packet,garncarek_parallel}. Related work also studies distributed scheduling in message-passing models \cite{georgiou_distributed}.

Our setting differs because packets contend through randomized transmissions on shared multiple-access channels rather than being assigned to machines or scheduled centrally on links. Fast-channel evictions also require controlling switching costs between the fast channel and a reliable slow channel.

\section{Algorithm Specification}\label{sec:algorithm-overview}

We describe \mainAlgorithm\ (\mainAlgorithmAcrynom) and the intuition behind its design.

\subsection{Preliminaries}\label{sec:primitive}

We use separate procedures for population estimation, contention resolution, and channel selection.\medskip

\noindent{\bf Number of active packet estimates.}
\mainAlgorithmAcrynom calls a decentralized subroutine \estimateFunction\ when all surviving packets are on the slow channel. Its initial call returns a common estimate {\boldmath{$\widehat{n}$}} of $n$; later calls return a common estimate {\boldmath{$\widehat{n}_{\mathrm{cur}}$}} of the active population, {\boldmath{$\curN$}}. With high probability, both are within a constant factor of the population. Each later call takes {\boldmath{$\tau_{\mathrm{est}}(\widehat{n})$}} $=O(\log^2 \widehat{n})$ slots. Population-size approximation is standard \cite{jurdzinski2002energy}; Appendix~\ref{sec:estimate-n} specifies the guarantees used by \mainAlgorithmAcrynom.\medskip

\noindent{\bf \tstb contention primitive.}
\mainAlgorithmAcrynom uses \TSTB (\tstb) as its contention-resolution subroutine \cite{bender2016contention}. A fresh execution with size parameter $z$ follows a predetermined sequence of contention windows. In each window, every remaining participant chooses one transmission opportunity uniformly and independently. Successful packets leave. The prescribed \defn{completion window} contains $L(z)=O(z)$ opportunities, including $2z$ in the first window, so its duration at rate $\nu\in\{1,\rho\}$ is $T_\nu(z)=\Theta(z/\nu)$ when $z\geq\nu$.

We rely on the following properties of \tstb. For a population within a fixed constant factor of an appropriate size parameter, a constant fraction succeeds except with probability $\exp(-\Omega(z))$. If at most $z$ packets participate in a fresh execution with size parameter $z$, all succeed except with probability $O(z^{-d})$, for a fixed $d\geq2$. Executions can be paused by stopping their common logical clock; permanent departures do not delay the remaining packets under the fixed schedule. Further details of the execution and its analysis appear in Appendix~\ref{sec:execution-details}.

\begin{figure}[t]
\centering
\noindent\rule{\linewidth}{0.4pt}
\begin{minipage}{\linewidth}
\RestyleAlgo{plain}
\begin{algorithm}[H]
\small
Run the initial \estimateFunction\ call to obtain $\widehat{n}$; set $\widehat{n}_{\mathrm{cur}}\gets \widehat{n}$, $b\gets0$, and $\tau_{\mathrm{pop}},\tau_{\mathrm{time}}$ as in Equation~\eqref{eq:two-cutoffs}\;\label{algo:line-one}
Begin an estimation epoch: set $W\gets0$ and initialize a paused slow \tstb instance with parameter $s\gets2\widehat{n}_{\mathrm{cur}}$\;\label{algo:begin-epoch}
\While{this packet remains active}{
 \If{$t\geq \tau_{\mathrm{time}}$\label{algo:time-cutoff}}{Enter permanent slow-channel cleanup with guess $\widehat{n}$\;\label{algo:time-cleanup}}
 \ElseIf{$\widehat{n}_{\mathrm{cur}}\leq\max\{\tau_{\mathrm{pop}},2\lambda\rho\}$\label{algo:population-cutoff}}{Enter permanent slow-channel cleanup with guess $\widehat{n}_{\mathrm{cur}}$\;}\label{algo:enter-permanent-slow}
 \ElseIf{$W\geq\delta \widehat{n}_{\mathrm{cur}}$\label{algo:second-else-if}}{
   Run \estimateFunction\ with bound $\widehat{n}$, checking $\tau_{\mathrm{time}}$ at every slow boundary\;
   If the call finishes first, update $\widehat{n}_{\mathrm{cur}}$ and begin a new estimation epoch, retaining $b$\;\label{algo:call-finish-first}
 }
 \ElseIf{fast availability is observed\label{algo:third-else-if}}{
   Set $x\gets\min\{\rho+b,\lfloor \widehat{n}_{\mathrm{cur}}/\lambda\rfloor\}$; choose a fast subchannel or remain on the slow channel using the \emph{Subchannel Assignment} rule (described below)\;
   Execute \tstb on each fast subchannel for $D(x)$ slots; any packets evicted before the end that have not succeeded return to the slow channel and wait\;\label{algo:execute-tstb}
   Return survivors at the deadline; use one settling slot and one reporting slot\;
   \eIf{the report is nonempty}{Set $b\gets0$\;}{Set $b\gets b+1$ and $W\gets W+\lambda x$\;\label{algo:third-else-if-else}}
 }
 \Else{\nllabel{algo:slow-channel-else}
   Execute the next slot of the slow-channel \tstb execution; if a packet succeeds, set $W\gets W+1$\;
   If its completion window ends, start a fresh instance with parameter $2\widehat{n}_{\mathrm{cur}}$, retaining $W$\;\label{algo:completion-window-ends}
 }
}
\end{algorithm}
\noindent\rule{\linewidth}{0.4pt}
\end{minipage}
\caption{\mainAlgorithm\ (\mainAlgorithmAcrynom)}\label{fig:race}\label{alg:channel-switch}\label{our-algorithm}
\end{figure}

\subsection{Algorithm Specification and Intuition}\label{sec:protocol}\label{sec:alg-intuition}

\mainAlgorithmAcrynom seeks constant-competitive latency while limiting switching cost. The slow channel provides progress during evictions; gradually increasing fast batches exploit availability without exposing too many packets to eviction.\medskip

\begin{figure}[t]
    \centering
    \includegraphics[width=\linewidth]{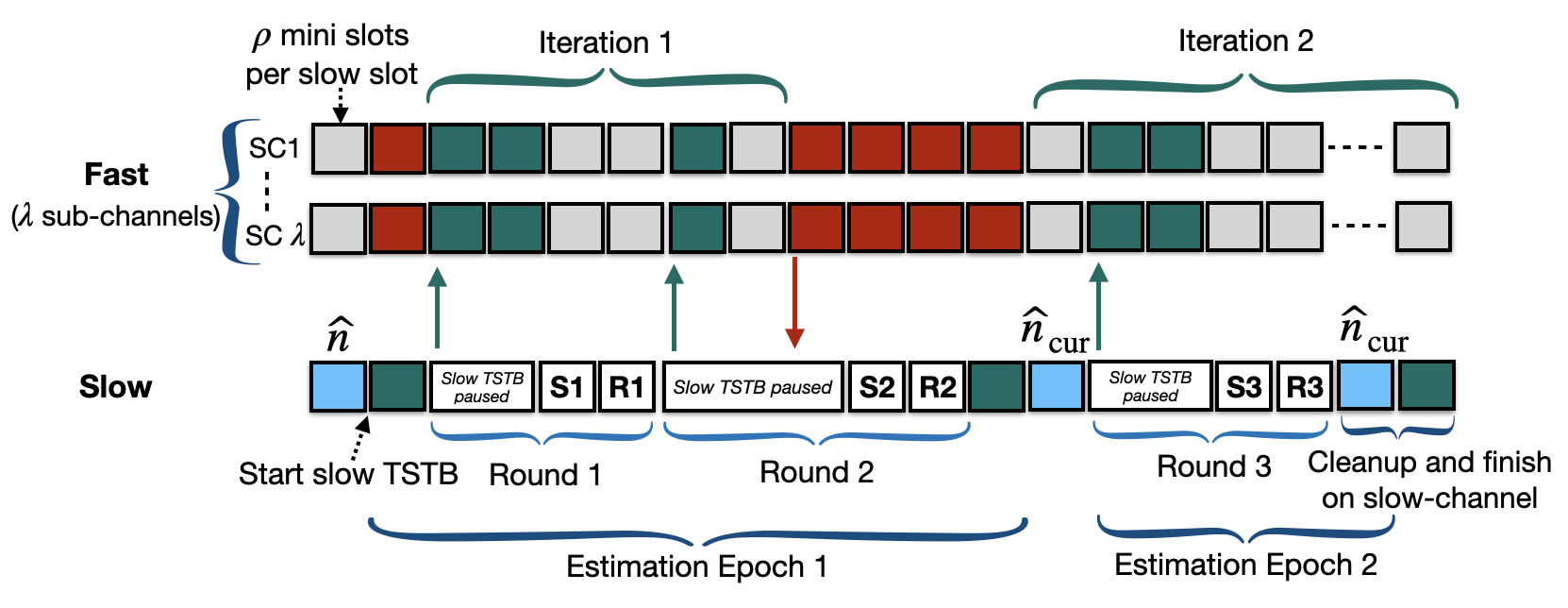}
   \caption{High-level illustration of \mainAlgorithmAcrynom.
On the fast channel, each box represents $\rho$ mini-slots. Teal boxes, on the fast-channel and slow channel, indicate activity, which may include successful transmissions, collisions, and empty mini-slots. Gray boxes represent idle slots, dark-red boxes represent fast-channel eviction periods, and blue boxes represent executions of \estimateFunction, each spanning multiple slow-channel slots.
Upward arrows indicate batches moving onto the fast channel, while downward arrows indicate packets returning to the slow channel after eviction. The initial population estimate is $\widehat n$, and subsequent estimates are denoted by $\widehat n_{\mathrm{cur}}$; each becomes available when its estimation finishes.
An iteration is a maximal interval of fast-channel availability. Each fast round includes \tstb execution, a settling slot ($S_i$), and a reporting slot ($R_i$). An estimation epoch begins when a population estimate is obtained and ends when the next estimation begins or permanent slow-channel cleanup starts. The slow-channel \tstb execution is paused during fast rounds and resumes afterward. Once a cutoff is reached, \mainAlgorithmAcrynom finishes the remaining packets on the slow channel.}
\label{fig:race-overview}
\end{figure}

\noindent{\bf Shared State and Specification.}
Figure~\ref{alg:channel-switch} gives the pseudocode, and Figure~\ref{fig:race-overview} provides a high-level illustration for \mainAlgorithmAcrynom . All surviving packets maintain the same fast-batch index $b$, current population estimate $\widehat{n}_{\mathrm{cur}}$, and \defn{completed-work counter} $W$. Following the initial call to \estimateFunction\ (lines~\ref{algo:line-one}--\ref{algo:begin-epoch}), $\widehat{n}_{\mathrm{cur}}=\widehat{n}$ and $b=W=0$.

The counter $W$ tracks progress since the most recent estimate and determines when to update it. It is not an exact count of successful packets: slow-channel \tstb successes are counted directly, while progress on the fast channel is credited using intended batch sizes, as specified below. A sufficiently small fixed constant $\delta>0$ sets the update threshold $W\geq\delta\widehat{n}_{\mathrm{cur}}$.

\mainAlgorithmAcrynom uses two fixed cutoffs to determine when it enters \defn{permanent slow-channel cleanup}, a final stage in which it no longer admits fast batches:
\begin{equation}\label{eq:two-cutoffs}
 \tau_{\mathrm{pop}}=\left\lceil c_{\mathrm{pop}}\,\widehat{n}/\rho\right\rceil,
 \qquad \tau_{\mathrm{time}}=\max\{1,\lfloor\widehat{n}/64\rfloor\}.
\end{equation}
The \defn{population cutoff} $\tau_{\mathrm{pop}}$ causes \mainAlgorithmAcrynom to stop using the fast channel once its estimate of the remaining population is sufficiently small. The test in lines~\ref{algo:population-cutoff}--\ref{algo:enter-permanent-slow} also includes the safeguard $2\lambda\rho$, so cleanup begins when $\widehat{n}_{\mathrm{cur}}\leq\max\{\tau_{\mathrm{pop}},2\lambda\rho\}$. The \defn{time cutoff} $\tau_{\mathrm{time}}$ stops new fast-channel admissions once that many slow slots have elapsed, regardless of the remaining population (lines~\ref{algo:time-cutoff}--\ref{algo:time-cleanup}). Time is measured from the beginning of execution, including the initial estimation call. A fast round already underway finishes before cleanup begins. With high probability, $n\leq\widehat{n}\leq32n$, so $\tau_{\mathrm{pop}}=\Theta(n/\rho)$ and $n/128\leq\tau_{\mathrm{time}}\leq n/2$ for sufficiently large $n$. The constant $c_{\mathrm{pop}}$ is chosen sufficiently large relative to the augmentation and estimation constants.\medskip

\noindent{\bf How Cleanup Works.} During cleanup, all remaining packets run \tstb on the slow channel, doubling the size parameter after each unsuccessful execution until all packets succeed. The initial parameter is based on $\widehat{n}_{\mathrm{cur}}$ for the population cutoff and $\widehat{n}$ for the time cutoff, with a fixed minimum if the estimate is too small. Appendix~\ref{sec:execution-details} specifies the procedure; Lemma~\ref{lem:cleanup} bounds its expected completion time.

After each population estimate, \mainAlgorithmAcrynom resets $W=0$ and initializes a slow-channel \tstb execution with parameter $2\widehat{n}_{\mathrm{cur}}$ (line~\ref{algo:begin-epoch}, and lines~\ref{algo:second-else-if}--\ref{algo:call-finish-first} for later estimates). An \defn{estimation epoch} begins when a population estimate is obtained and ends when the next call to \estimateFunction\ begins or permanent slow-channel cleanup starts. Updating the estimate does not reset $b$.

When the fast channel is available, \mainAlgorithmAcrynom admits one batch of packets (lines~\ref{algo:third-else-if}--\ref{algo:third-else-if-else}). It partitions this batch among the $\lambda$ logical subchannels, with an intended number of $x$ packets per subchannel, where:
$$
 x=\min\{\rho+b,\lfloor\widehat{n}_{\mathrm{cur}}/\lambda\rfloor\}.
$$
The intended total batch size is $\lambda x$; although, since packets independently choose whether to join and which subchannel to use, the actual number selected may differ. Each subgroup executes a fresh \tstb instance with parameter $2x$ at rate $\rho$. All surviving packets use the same scheduled duration:
$$
 D(x)=\lceil L(2x)/\rho\rceil=\Theta(x/\rho).
$$

We call the operation of admitting and processing this batch, followed by two slots to check for an eviction, a \defn{fast round}, or simply a \defn{round}. Even if every selected packet succeeds early, \mainAlgorithmAcrynom waits until the $D(x)$ slots have elapsed. Any packets still on the fast channel then return to the slow channel. During the first of the two additional slow slots, no packet transmits. During the second, packets that were evicted before the scheduled end and have not yet succeeded transmit to report the eviction (lines~\ref{algo:execute-tstb}--\ref{algo:third-else-if-else}). If that slot is nonempty, whether a success or a collision, makes the round \defn{aborted}: \mainAlgorithmAcrynom resets $b=0$ and leaves $W$ unchanged, although some fast transmissions may have succeeded. Otherwise the round is \defn{certified}: \mainAlgorithmAcrynom increments $b$ and adds $\lambda x$ to $W$.  Thus, apart from the cap in the definition of $x$, the intended number of packets per subchannel rises from $\rho$ to $\rho+1$, $\rho+2$, and so on, until a reported abort resets it to $\rho$. \mainAlgorithmAcrynom admits no new batch until the round is over.

When the fast channel is unavailable, \mainAlgorithmAcrynom runs the next slot of slow-channel \tstb (lines~\ref{algo:slow-channel-else}--\ref{algo:completion-window-ends}), adding one to $W$ for each success. This execution pauses during fast rounds and resumes afterward. If it reaches the end of its prescribed windows while packets remain, a new execution begins with parameter $2\widehat{n}_{\mathrm{cur}}$, retaining $W$. The algorithm checks its conditions again at the next slow-slot boundary.

At each slow-slot boundary or after a fast round, \mainAlgorithmAcrynom checks the cutoffs, then the estimation threshold (lines~4--10). When $W\geq\delta\widehat{n}_{\mathrm{cur}}$, it runs \estimateFunction\ on the slow channel. If estimation finishes, it resets $W=0$ and begins a new estimation epoch without resetting $b$. Fast-channel availability does not interrupt estimation, but reaching the time cutoff sends the algorithm to cleanup.\medskip

\noindent{\bf Role of the Two Cutoffs.} The cutoffs preserve constant-competitive latency while ensuring that \mainAlgorithmAcrynom stops admitting fast batches by the time \optimalAlgo finishes, with high probability. The \emph{population cutoff} triggers cleanup when only $O(n/\rho)$ packets remain; these can finish on the slow channel in expected $O(n/\rho)$ time.

The \emph{time cutoff} handles later completion by \optimalAlgo: when \optimalAlgo needs $\Omega(n)$ slots, slow-channel cleanup from any remaining population is affordable. If \optimalAlgo finishes earlier, our analysis shows that augmented fast-channel capacity forces \mainAlgorithmAcrynom to reach the population cutoff in time. Thus neither case requires additional fast admissions after \optimalAlgo finishes.\medskip

\noindent{\bf Subchannel Assignment.}
In each fast round, every active packet independently chooses subchannel $j$ with probability $x/\widehat{n}_{\mathrm{cur}}$ for each $j\in\{1,\ldots,\lambda\}$, or remains on the slow channel otherwise. This determines both batch membership and subchannel assignment; the numbers selected need not be equal. All selected packets enter as one batch and use only their assigned subchannel. Packets remaining on the slow channel pause \tstb until the round ends, and unsuccessful packets returning from the fast channel resume their previous slow-channel execution.\medskip

\begin{figure}[t]
\centering
\includegraphics[width=0.83\linewidth]{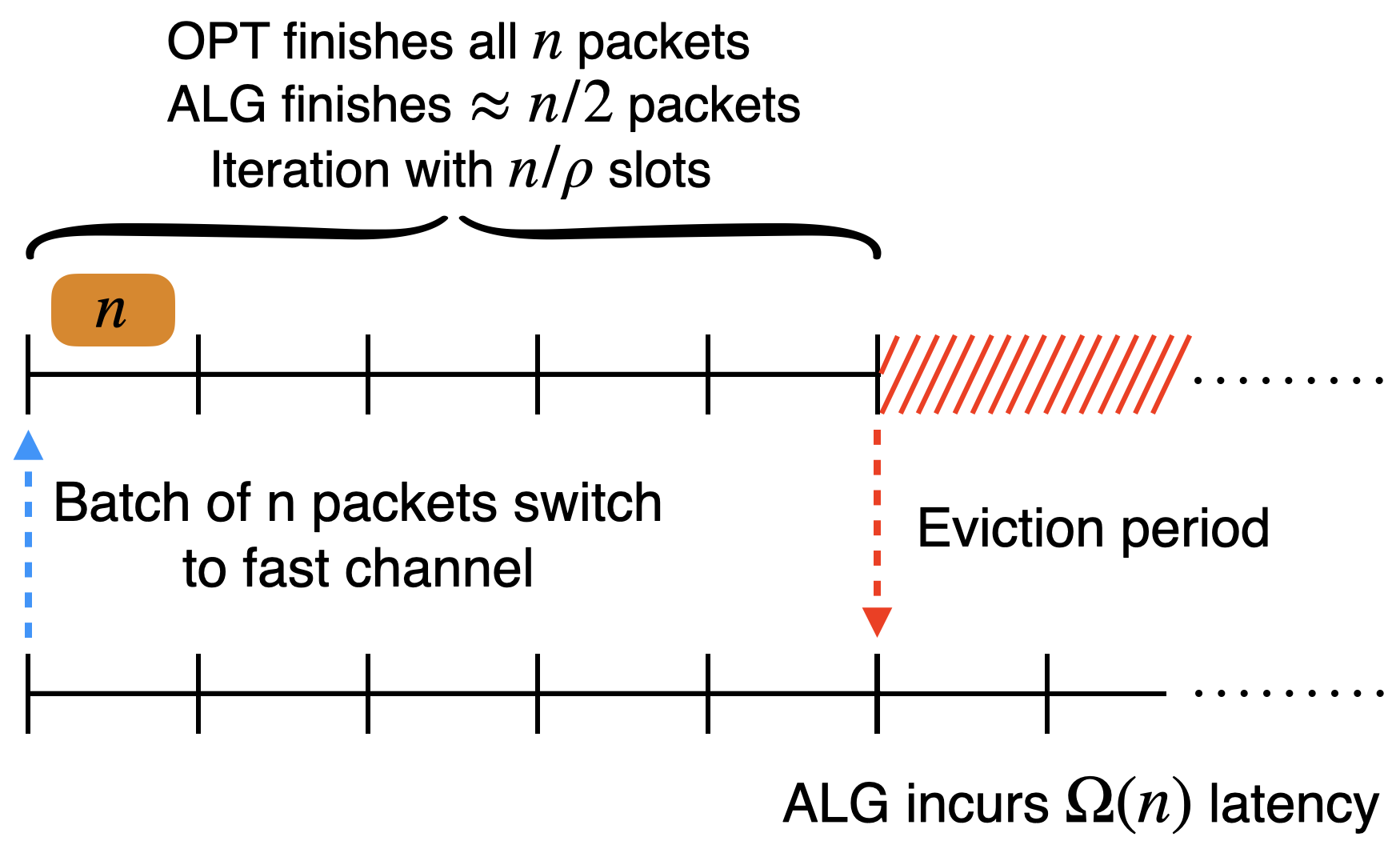}
\caption{Illustrating the need for fast-channel augmentation. During an initial $\Theta(n/\rho)$-slot interval, \optimalAlgo can complete all $n$ packets on the fast channel, while an unaugmented algorithm, ALG, may complete only a constant fraction. If the fast channel then becomes permanently unavailable, ALG requires $\Omega(n)$ additional slots to finish on the slow channel.}
\label{fig:fast-augmentation-needed}
\end{figure}

\noindent{\bf Motivation for Gradually Increasing Batch Sizes.} Each fast batch costs one admission, while an eviction can return many unfinished packets. Small batches limit eviction cost but increase admissions; large batches do the reverse. \mainAlgorithmAcrynom raises the intended size per subchannel by one after a certified round and resets it to $\rho$ after an abort. Reaching size $x>2\rho$ requires certifying sizes $\rho,\rho+1,\ldots,x-1$, which complete $\Omega(x^2)$ packets with high probability. The next eviction exposes only $O(x)$ packets per subchannel. Thus, larger risks follow substantial progress. \medskip

\noindent{\bf Resource Augmentation.} \optimalAlgo knows future evictions and schedules without collisions. The extra fast-channel capacity of \mainAlgorithmAcrynom compensates for these advantages. A wireless interpretation is additional bandwidth divided into the logical subchannels described in Section~\ref{sec:our-model}. Figure~\ref{fig:fast-augmentation-needed} shows the difficulty without augmentation: in an initial interval of $n/\rho$ slots, \optimalAlgo can finish all packets, while an unaugmented contention-resolution algorithm may leave $\Theta(n)$ unfinished. If the fast channel then becomes unavailable indefinitely, slow-channel completion takes $\Omega(n)$ further slots, a factor of $\Theta(\rho)$ more than \optimalAlgo. Sufficient constant-factor augmentation on the fast channel prevents this loss.

The slow channel guarantees eventual delivery, but need not match \optimalAlgo's throughput at every moment. The time cutoff limits fast admissions and the population cutoff makes earlier slow-channel cleanup affordable. Together they remove the need for slow-channel augmentation.

\section{Technical Overview}\label{sec:tech-overview}

We outline the main technical challenges; full proofs appear in the appendix. Latency and switching cost pull in opposite directions: \mainAlgorithmAcrynom must exploit fast-channel availability without knowing the next eviction, but larger batches expose more packets to costly returns. Gradually increasing batch sizes balance these demands. The analysis must also maintain accurate population estimates and prove that fast admissions stop by the time \optimalAlgo finishes. Otherwise, later fast-channel availability could lead to further switching costs even after \optimalAlgo has finished.

Throughout this overview, let $G$ denote the \emph{good event} that the initial and subsequent population estimates satisfy their guarantees and that the packet-assignment and progress guarantees of Lemma~\ref{lem:exp-number-revised} hold whenever its assumptions are met. Lemma~\ref{lem:good} gives probability $1-O(n^{-D})$ for any desired fixed $D$, by choosing constants appropriately. Cleanup and cutoff safeguards separately control the contribution of the complementary event.\medskip

\noindent{\bf Batch-Level Estimates.} Each fast subchannel has intended batch size $x$; a packet joins it with probability $x/\widehat{n}_{\mathrm{cur}}$. When this estimate is within a constant factor of $\curN$, the expected subgroup size is $\Theta(x)$. Since $x\geq\rho=\Omega(\log n)$, Chernoff bounds give $\Theta(x)$ participants except with probability $\exp(-\Omega(x))$. Fresh \tstb then yields $\Omega(x)$ successes with the same exponential failure bound. Lemma~\ref{lem:exp-number-revised} establishes this simultaneously across subchannels: every certified round of target $x$ completes at least $\alpha\lambda x$ packets on $G$, for a constant $\alpha>0$ independent of $\lambda$.

When a fast-channel batch completes its scheduled \tstb execution without interruption, the expected number of unsuccessful packets returning to the slow channel is $O(1)$ per subchannel, by the \tstb guarantee and the concentration bound on subgroup sizes. Lemma~\ref{lem:non-evicted-batch-cost} therefore establishes an expected return cost of $O(\lambda)=O(1)$ for the entire batch.\medskip

\noindent{\bf Tracking Progress Across Channels.} The analysis of the counter $W$ requires two properties: (i) the population estimate must remain accurate until $W$ triggers an update, and (ii) each update must correspond to a substantial decrease in the active number of packets. The counter can differ from the actual progress in either direction: a certified round credits $\lambda x$ even if fewer packets succeed, while an aborted round may complete packets without increasing $W$.\smallskip

\noindent{\it Maintaining Estimate Accuracy.}
The main challenge is that aborted rounds may have completed
many packets without increasing $W$. We charge those possible departures to preceding certified work. An aborted round with intended size $\rho+t$, $t\geq1$, follows certified rounds of intended sizes $\rho,\ldots,\rho+t-1$ since the last reset of $b$. Since:
$$
 \rho+t\leq2\sum_{i=0}^{t-1}(\rho+i),
$$
the number of packets that could have succeeded during the aborted round is at most a constant multiple of the total intended size of the preceding certified batches.

A complication arises when a round with $x=\rho$ is aborted, since there may be no preceding certified rounds to account for it. However, the minimum-availability condition (recall $C_{\mathrm{avail}}$ from Section \ref{sec:our-model}) prevents two such rounds from both being aborted without an intervening certified round or population estimate. Moreover, an aborted round with $x=\rho$ that follows an aborted round with $x>\rho$ can be charged to the earlier abort. Lemma~\ref{lem:charge} shows that the sum of $x$ over aborted rounds is at most a constant multiple of the sum of $x$ over certified rounds, with an additional $O(\rho)$ for initialization and each population estimate.

For a prefix of estimation epoch $j$, let $C_s$ count slow-channel \tstb successes, and let $V_j$ and $I_j$ be the sums of the intended  batch sizes $x$ on the fast channel over certified and aborted rounds, respectively. Then $W=C_s+\lambda V_j$. On $G$, every intended per-subchannel batch size is at most:
$$
F=O\left(\rho+\sqrt{n/\lambda}\right)=O(\sqrt n).
$$
Within the epoch, Lemma~\ref{lem:charge} gives $I_j\leq4V_j+4F$. The additive term accounts for an aborted round whose preceding certified rounds may belong to an earlier estimation epoch.

On event $G$, each fast round contains at most $2\lambda x$ packets, regardless of whether it is certified or aborted. Using Lemma~\ref{lem:charge}, we can therefore bound the number of packets that have succeeded since the last population estimate by:
$$
C_s+2\lambda(V_j+I_j)+2\lambda F
\leq 10W+10\lambda F.
$$
Let $\widehat{n}_{\mathrm{cur},j}$ denote the estimate at the beginning of estimation epoch $j$, and let $N_j$ be the actual number of active packets at that time. The bound above, together with the rule for updating $W$, shows that sufficiently small $\delta$ and sufficiently large $c_{\mathrm{pop}}$ ensure that only a small fraction of these packets can succeed before the next estimate is obtained. Consequently, the estimate remains within a constant factor of the actual number of active packets throughout the epoch, as established in Lemma~\ref{lem:invariant}.\smallskip

\noindent{\it Ensuring Substantial Progress.}
When $W$ reaches $\delta\widehat{n}_{\mathrm{cur},j}$,
\mainAlgorithmAcrynom specifies that a new population estimate
is performed. Every slow-channel \tstb success contributes one to $W$, while each certified fast round completes at least an $\alpha$ fraction of the number credited to $W$. Consequently, the number of packets that have succeeded since the beginning of the epoch is at least $\alpha W\geq\alpha\delta N_j$.

Hence, if another estimation epoch follows, the number of active packets satisfies $N_{j+1}\leq(1-\alpha\delta)N_j$. Since $\alpha\delta>0$ is a constant, the number of active packets decreases by a constant factor between successive epochs, so \mainAlgorithmAcrynom invokes the estimator only $O(\log n)$ times. Moreover, because each estimate is within a constant factor of the actual number of active packets, the sum of these estimates over all epochs is $O(n)$.\medskip

\noindent{\bf Progress on the Fast-Channel.}
On event $G$, \mainAlgorithmAcrynom invokes \estimateFunction\ at most $O(\log n)$ times, and each invocation takes $O(\log^2 n)$ slots. Including the initial estimate, the total time spent estimating is therefore $O(\log^3 n)$. During these slots, \mainAlgorithmAcrynom does not use the fast channel, even if it is available. Since \optimalAlgo can transmit at most $\rho$ packets per slot on the fast channel, \mainAlgorithmAcrynom misses at most $O(\rho\log^3 n)=o(n)$ fast-channel transmission opportunities due to estimation. Thus, the total loss from estimation is small, even if some fast-channel availability intervals are spent entirely on estimating the number of active packets.

Let $\Service_f([0,t))$ denote the total number of fast-channel mini-slots during which the fast channel is available before time $t$. Since \optimalAlgo can transmit at most one packet per mini-slot on the fast channel and one packet per slow-channel slot, the total number of packets \optimalAlgo can complete by time $t$ is at most:
$$
\Service([0,t))=t+\Service_f([0,t)).
$$
This bound may count some fast-channel mini-slots that \optimalAlgo cannot use because packets can switch channels only at slow-slot boundaries, but this only makes the upper bound larger. Lemma~\ref{lem:fast-budget} shows that, while \mainAlgorithmAcrynom has not completed or entered cleanup and $t<\tau_{\mathrm{time}}$, we have:
\begin{equation}\label{eq:overview-service}
\Service([0,t))\leq t+O(n/\lambda)+O(\rho\log^3 n+\sqrt n).
\end{equation}
This holds on event $G$, where the constant implicit in $O(n/\lambda)$ is independent of $\lambda$. We establish this bound by accounting for three situations in which the fast channel may be available to \optimalAlgo: while \mainAlgorithmAcrynom executes fast rounds, while \mainAlgorithmAcrynom estimates the number of active packets, and when the fast channel becomes available partway through a slow-channel slot.\medskip

\noindent{\it During fast rounds.}
To bound the number of fast-channel transmission opportunities available to \optimalAlgo during \mainAlgorithmAcrynom's fast rounds, we first bound the sum of the intended per-subchannel batch sizes $x$ over those rounds.
Let $V$ and $I$ be the sums of the intended per-subchannel batch sizes $x$ over certified and aborted rounds, respectively, completed by time $t$. Since certified rounds complete at least $\alpha\lambda V$ packets on event $G$, we have
$V\leq\frac{n}{\alpha\lambda}$.

Let $h$ be the number of population estimates initiated by time $t$. Lemma~\ref{lem:charge} establishes that  $I\leq4V+\rho(h+1)$. Accounting also for a round that may still be in progress at time $t$, whose intended per-subchannel batch size is at most $F$, we obtain:
$$
\sum_{\text{rounds started by }t}\hspace{-25pt}x
~~~\leq V+I+F
\leq\frac{5n}{\alpha\lambda}+\rho(h+1)+F.
$$
Recall from specification of \mainAlgorithmAcrynom (Section \ref{sec:alg-intuition}) that each fast round lasts $D(x)+2=O(x/\rho)$ slow slots, including the two additional slots used to determine whether an eviction occurred. During this time, \optimalAlgo has at most $O(x)$ fast-channel transmission opportunities. Thus, the bound on the sum of $x$ above shows that the total number of such opportunities during \mainAlgorithmAcrynom's fast rounds is at most:
$$
O(n/\lambda)+O(\rho h+F).
$$
The factor $1/\lambda$ arises because \mainAlgorithmAcrynom makes progress on $\lambda$ logical subchannels simultaneously, whereas \optimalAlgo has only one fast channel.\medskip

\noindent{\it During population estimation.} As established above, \mainAlgorithmAcrynom spends a total of $O(\log^3 n)$ slots estimating the number of active packets, including the initial estimate. \mainAlgorithmAcrynom does not use the fast channel during these slots, even if it is available. Consequently, \optimalAlgo has at most $O(\rho\log^3 n)$ fast-channel transmission opportunities during these periods.\medskip

\noindent{\it When the fast channel becomes available during a slow-channel slot.}
\mainAlgorithmAcrynom checks whether the fast channel is available only at slow-slot boundaries. If the channel becomes available partway through a slot, \mainAlgorithmAcrynom may therefore miss up to $\rho$ fast-channel transmission opportunities before it can respond. The minimum-availability condition ensures that the fast channel remains available at the next boundary, when \mainAlgorithmAcrynom begins either a fast round or a population estimate. We can account for at most one such delay before each of these operations. For fast rounds, this additional cost is absorbed into the bound above because $x\geq\rho$. The delays preceding population estimates contribute at most $O(\rho(h+1))$ additional opportunities.

Combining these three bounds, using $h=O(\log n)$ and $F=O(\sqrt n)$, gives the fast-channel contribution in Equation~\ref{eq:overview-service}.\medskip

\noindent{\it Why \mainAlgorithmAcrynom stops using the fast channel when \optimalAlgo finishes early.} Finally, \optimalAlgo can complete at most $t$ packets on the slow channel by time $t$, regardless of \mainAlgorithmAcrynom's progress on that channel. Our analysis does not require \mainAlgorithmAcrynom to match \optimalAlgo's slow-channel throughput. By choosing the fast-channel augmentation factor $\lambda$ sufficiently large, the $O(n/\lambda)$ term can be made at most $n/8$. Since $\rho=O(\sqrt n)$, the remaining lower-order terms contribute at most another $n/8$ for sufficiently large $n$. Consequently, as long as \mainAlgorithmAcrynom has not completed or entered cleanup and $t<\tau_{\mathrm{time}}$, we have:
$$
\Service([0,t))\leq t+n/4.
$$

\noindent{\bf Combining the Bounds.} We now use the bound on \optimalAlgo's available service to establish our latency guarantee and show that \mainAlgorithmAcrynom stops admitting new fast batches by the time \optimalAlgo finishes. Let $T^*=\LatRV(\optimalAlgo,I)$ denote \optimalAlgo's completion time. We consider two cases, depending on whether \optimalAlgo finishes before or after \mainAlgorithmAcrynom's time cutoff. \mainAlgorithmAcrynom itself does not need to distinguish between these cases.\smallskip

\noindent{\it Case 1: \optimalAlgo finishes before the time cutoff.}
Suppose $T^*<\tau_{\mathrm{time}}\leq n/2$. Since \optimalAlgo completes all $n$ packets by time $T^*$, we must have $n\leq\Service([0,T^*))$. However, if \mainAlgorithmAcrynom had neither completed nor entered cleanup by $T^*$, the preceding bound would imply:
$$
n\leq\Service([0,T^*))\leq T^*+n/4<n,
$$
which is a contradiction. Therefore, \mainAlgorithmAcrynom must have completed or entered permanent slow-channel cleanup by the time \optimalAlgo finishes. Since the time cutoff has not been reached, this cleanup is triggered by the population cutoff, leaving only $O(n/\rho)$ packets to finish. Their expected completion time is $O(n/\rho)=O(T^*)$. Moreover, \mainAlgorithmAcrynom admits no further fast batches after \optimalAlgo finishes, even if the fast channel subsequently becomes available.\smallskip

\noindent{\it Case 2: \optimalAlgo finishes at or after the time cutoff.}
If $T^*\geq\tau_{\mathrm{time}}$, then $T^*\geq n/128=\Omega(n)$ on the initialization good event. The time cutoff prevents further fast admissions, and even slow-only cleanup from all $n$ packets costs expected $O(n)=O(T^*)$. A round already admitted before $\tau_{\mathrm{time}}$ finishes normally and adds at most $O(n)$ time; its return cost is charged below. Thus this case requires no slow-channel progress advantage.

This explains why a larger fixed population threshold is not an interchangeable substitute for $\tau_{\mathrm{time}}$. Cleaning up $\Theta(n)$ packets could be too slow when \optimalAlgo finishes in $\Theta(n/\rho)$ time, whereas it is affordable in the second case. The population cutoff handles the first case, and the time cutoff handles the second. Lemma~\ref{lem:total-progress-to-opt-time} formalizes the absence of new fast admissions after \optimalAlgo, and Proposition~\ref{lem:latency-constant-factor} establishes expected constant-competitive latency, including exceptional estimates.\medskip

\noindent{\bf Switching Cost.}
For the switching-cost comparison, assume $\TNI\geq1$. Let $J$ be the number of reported aborts, and let $t_1,\ldots,t_{J+1}$ count the certified rounds before the first abort, between successive aborts, and after the last abort. Each abort resets $b$, whereas a new population estimate does not. Thus, the certified rounds counted by $t_i$ have intended per-subchannel batch sizes $\rho,\rho+1,\ldots,\rho+t_i-1$. On event $G$, each certified round with intended size $x$ completes at least $\alpha\lambda x$ packets. Since each packet succeeds only once, we have:
$$
\frac{\alpha\lambda}{2}\sum_{i=1}^{J+1}t_i^2\leq n.
$$

On event $G$, \mainAlgorithmAcrynom begins no new fast rounds after \optimalAlgo finishes. Each reported abort results from an eviction, so two aborted rounds cannot have begun during the same period of fast-channel availability. Lemma~\ref{lem:offline} bounds the number of such periods before \optimalAlgo finishes by $\TNI+2$. Hence $J\leq\TNI+2$, and Cauchy--Schwarz gives:
$$
\sum_{i=1}^{J+1}t_i
\leq\sqrt{(J+1)\sum_{i=1}^{J+1}t_i^2}
=O(\sqrt{\TNI n}).
$$

There are at most $\sum_i t_i+J+1$ fast rounds, each incurring one batch-admission cost. An aborted round following $t_i$ certified rounds has intended per-subchannel size $\rho+t_i$ and, on $G$, can return at most $2\lambda(\rho+t_i)$ packets through eviction. Thus, the total cost of admissions and returns caused by evictions is:
$$
O\left((J+1)\rho+\sum_{i=1}^{J+1}t_i\right)
=O(\TNI\rho+\sqrt{\TNI n}),
$$
as established in Lemma~\ref{lem:large-iterations}.

We must also account for packets that return to the slow channel because they have not succeeded by the scheduled end of a fast round. Lemma~\ref{lem:return} shows that the expected number of such packets is $O(\lambda)=O(1)$ per round, provided the estimate of the number of active packets is sufficiently accurate. Since \mainAlgorithmAcrynom begins at most $O(\TNI\rho+\sqrt{\TNI n})$ rounds on event $G$, Lemma~\ref{lem:expectedcost} shows that these returns contribute at most $O(\TNI\rho+\sqrt{\TNI n})$ to the expected switching cost. This includes packets returning after \optimalAlgo finishes, provided their fast round began before \optimalAlgo's completion.

The contribution from executions outside event $G$ is negligible, as established using the time cutoff and the guarantees of \estimateFunction. Combining all switching costs gives:
$$
E[\Cost(\mainAlgorithmAcrynom,I)]=O(\TNI\rho+\sqrt{\TNI n}).
$$
Since \optimalAlgo incurs at least one switching cost in each of the $\TNI$ fast-channel availability intervals it uses, the expected switching-cost competitive ratio is
$O(\rho+\sqrt{n/\TNI})$. Together with the latency bound, this proves Theorem~\ref{thm:main-theorem}.\medskip

\noindent{\bf Lower Bounds.} For  Theorem~\ref{thm:lower-bounds}(A), we choose no eviction with probability \(1/2\), or otherwise introduce a one-mini-slot eviction at a uniformly random fast mini-slot boundary strictly inside a slow-channel slot, after the required initial availability period. \optimalAlgo finishes in $O(n/\rho)$ slots with constant switching cost. Even a one-mini-slot eviction can force the algorithm to wait until the next slow-slot boundary before returning to the fast channel.

By Markov's inequality, with constant probability over the algorithm's random choices, it finishes the no-eviction instance quickly. Fix any such choice of randomness and consider the resulting execution. Slow-channel successes and fast-channel successes in the first constant number of slow slots total only $o(n)$, so $\Omega(n)$ packets succeed on the fast channel thereafter, before the algorithm finishes. We group those packets by their last move onto the fast channel before success. If there are many groups, batch admissions already incur substantial switching cost. Otherwise, packets remain on the fast channel long enough that a randomly chosen eviction incurs substantial cost. For group sizes $s_i$, the total time spent by packets on the fast channel, summed over packets and mini-slots, is $\Omega(\sum_i s_i^2)$ after accounting for constant-size groups and excluding mini-slot boundaries that coincide with slow-slot boundaries. Cauchy--Schwarz gives us the tradeoff:
$$
\Omega(g)+\Omega(n/g)=\Omega(\sqrt n),
$$
where $g$ is the number of groups. Averaging over the algorithm's random choices and the distribution of eviction schedules shows that some fixed eviction schedule yields the claimed expected switching-cost lower bound.

The second lower bound considers an eviction that begins immediately after the first mini-slot of a slow-channel slot and leaves the fast channel permanently unavailable. The key observation is that, to achieve latency within a constant factor of \optimalAlgo, an online algorithm must have $\Omega(\rho)$ packets on the fast channel during a constant fraction of the relevant slow slots, with constant probability. By choosing the eviction time uniformly at random from these slots, independently of the algorithm's random choices, we obtain expected eviction cost $\Omega(\rho)$. Averaging gives a fixed schedule with expected switching-cost ratio $\Omega(\rho)$, since \optimalAlgo attains minimum latency with unit switching cost.

\section{Conclusion and Future Work}\label{sec:conclusion}

We introduced an online model for contention resolution under dynamic spectrum sharing, where a high-rate fast channel may be interrupted by evictions. This extends contention resolution to a setting where packets must balance latency against the cost of moving between channels. For $\Omega(\log n)\leq\rho\leq O(\sqrt n)$, \mainAlgorithmAcrynom achieves expected latency within a constant factor of \optimalAlgo and, when $\TNI\geq1$, expected switching-cost ratio $O(\rho+\sqrt{n/\TNI})$, using constant-factor augmentation on the fast channel alone. A small population cutoff and an early time cutoff allow the slow channel to remain unaugmented. Our lower bounds establish worst-case $\Omega(\sqrt n)$ ratio and a separate $\Omega(\rho)$ competitive-ratio lower bound; tight dependence on both $\rho$ and $\TNI$ remains open. Two further interesting directions are to extend the rate range and to allow dynamic arrivals. \medskip


\noindent\textbf{Acknowledgments and Generative AI Disclosure.}
This work was supported by the National Science Foundation (NSF) under Grant CNS-2210300. The research problem, model, algorithm design, and overall proof strategy are original contributions of the authors. ChatGPT was used to assist in simplifying and refining parts of the mathematical analysis, as well as improving the clarity and readability of the manuscript. The technical contributions and high-level exposition remain the work of the authors, who have reviewed all AI-assisted content and take full responsibility for the correctness of the results and any remaining errors.

\bibliographystyle{plain} 
\bibliography{CR}

@inproceedings{christodoulou2018strategic,
  title={Strategic Contention Resolution in Multiple Channels},
  author={Christodoulou, George and Melissourgos, Themistoklis and Spirakis, Paul G.},
  booktitle={Proceedings of the International Workshop on Approximation and Online Algorithms},
  pages={165--180},
  year={2018},
}

@inproceedings{jurdzinski_packet_jamming,
author = {Tomasz Jurdzi{\'{n}}ski and
Dariusz R. Kowalski and
Krzysztof Lory{\'{s}}},
title = {Online Packet Scheduling Under Adversarial Jamming},
booktitle = {Approximation and Online Algorithms - 12th International Workshop,
{WAOA} 2014, Wroc{\l}aw, Poland, September 11--12, 2014,
Revised Selected Papers},
series = {Lecture Notes in Computer Science},
volume = {8952},
pages = {193--206},
publisher = {Springer},
year = {2015},
doi = {10.1007/978-3-319-18263-6_17}
}

@article{chlebus2001randomized,
  title={Randomized communication in radio networks},
  author={Chlebus, Bogdan S.},
  journal={Handbook of Randomized Computing},
editor={Panos M. Pardalos and Sanguthevar Rajasekaran and  John H. Reif and Jose D. P. Rolim},
  volume={1},
  number={},
  pages={401--456},
  year={2001},
  publisher={Kluwer Academic Publishers}
}

@article{anta_speed_augmented_jamming,
author = {Antonio Fern{\'{a}}ndez Anta and
Chryssis Georgiou and
Dariusz R. Kowalski and
Elli Zavou},
title = {Competitive analysis of fundamental scheduling algorithms on a fault-prone
machine and the impact of resource augmentation},
journal = {Future Generation Computer Systems},
volume = {78},
pages = {245--256},
year = {2018},
doi = {10.1016/J.FUTURE.2016.05.042}
}

@article{georgiou_distributed,
author = {Chryssis Georgiou and
Dariusz R. Kowalski},
title = {On the competitiveness of scheduling dynamically injected tasks on processes prone to crashes and restarts},
journal = {Journal of Parallel and Distributed Computing},
volume = {84},
pages = {94--107},
year = {2015},
doi = {10.1016/j.jpdc.2015.07.007}
}

@inproceedings{garncarek_parallel,
author = {Pawe{\l} Garncarek and
Tomasz Jurdzi{\'{n}}ski and
Krzysztof Lory{\'{s}}},
title = {Fault-Tolerant Online Packet Scheduling on Parallel Channels},
booktitle = {2017 IEEE International Parallel and Distributed Processing Symposium (IPDPS)},
pages = {347--356},
publisher = {IEEE},
year = {2017},
doi = {10.1109/IPDPS.2017.105}
}

@ARTICLE{9383780,
  author={Aydin, Yucel and Kurt, Gunes Karabulut and Ozdemir, Enver and Yanikomeroglu, Halim},
  journal={IEEE Internet of Things Journal}, 
  title={Group Handover for Drone Base Stations}, 
  year={2021},
  volume={8},
  number={18},
  pages={13876-13887},
 }

@article{CAO20181,
title = {EGHR: Efficient group-based handover authentication protocols for mMTC in 5G wireless networks},
author={Cao, Jin and Ma, Maode and Li, Hui and Fu, Yulong and Liu, Xuefeng},
journal = {Journal of Network and Computer Applications},
volume = {102},
pages = {1-16},
year = {2018},
}

@article{omitola2018group,
  title={Group Handover Strategy for Mobile Relays in LTE-A Networks.},
  author={Omitola, Olusegun O and Srivastava, Viranjay M},
  journal={Journal of Communications},
  volume={13},
  number={9},
  pages={505--511},
  year={2018}
}

@article{xiao2022decentralized,
  title={Decentralized spectrum access system: Vision, challenges, and a blockchain solution},
  author={Xiao, Yang and Shi, Shanghao and Lou, Wenjing and Wang, Chonggang and Li, Xu and Zhang, Ning and Hou, Y Thomas and Reed, Jeffrey H},
  journal={IEEE Wireless Communications},
  volume={29},
  number={1},
  pages={220--228},
  year={2022},
  publisher={IEEE}
}

@article{englert2021online,
  title={Online makespan scheduling with job migration on uniform machines},
  author={Englert, Matthias and Mezlaf, David and Westermann, Matthias},
  journal={Algorithmica},
  volume={83},
  number={12},
  pages={3537--3566},
  year={2021},
  publisher={Springer}
}

@article{dosa2019tight,
  title={Tight lower bounds for semi-online scheduling on two uniform machines with known optimum},
  author={D{\'o}sa, Gy{\"o}rgy and F{\"u}genschuh, Armin and Tan, Zhiyi and Tuza, Zsolt and W{\k{e}}sek, Krzysztof},
  journal={Central European journal of operations research},
  volume={27},
  number={4},
  pages={1107--1130},
  year={2019},
  publisher={Springer}
}

@article{liu_online_two_uniform,
  author = {P. Liu and Xiwen Lu},
  title = {Online Scheduling of Two Uniform Machines to Minimize Total Completion Times},
  journal = {Journal of Industrial and Management Optimization},
  volume={5},
  number={1},
  pages={95-102},
  year = {2009}
}

@inproceedings{bender2005adversarial,
    Author = {Michael A. Bender and Martin Farach-Colton and Simai He and Bradley C. Kuszmaul and Charles E. Leiserson},
    Booktitle = {Proceedings of the $17$th Annual ACM Symposium on Parallelism in Algorithms and Architectures (SPAA)},
    Pages = {325-332},
    Title = {Adversarial Contention Resolution for Simple Channels},
    Year = {2005},
    }

@inproceedings{gereb1992efficient,
  title={Efficient optical communication in parallel computers},
  author={Ger{\'e}b-Graus, Mih{\'a}ly and Tsantilas, Thanasis},
  booktitle={Proceedings of the 4th Annual ACM Symposium on Parallel Algorithms and Architectures},
  pages={41--48},
  year={1992}
}

@inproceedings{greenberg1985randomized,
	Author = {Ronald I. Greenberg and Charles E. Leiserson},
	Bibnote = {Was Greenberg-Leiserson-1985},
	Booktitle = {Proceedings of the 26th Annual Symposium on the Foundations of Computer Science (FOCS)},
	Pages = {241-249},
	Title = {Randomized Routing on Fat-Trees},
	Year = {1985}}

@inproceedings{chlebus2007stability,
  title={Stability of the multiple-access channel under maximum broadcast loads},
  author={Chlebus, Bogdan S and Kowalski, Dariusz R and Rokicki, Mariusz A},
  booktitle={Proceedings of the Symposium on Self-Stabilizing Systems (SSS)},
  pages={124--138},
  year={2007},
  organization={Springer}
}

@inproceedings{anantharamu2010deterministic,
  title={Deterministic broadcast on multiple access channels},
  author={Anantharamu, Lakshmi and Chlebus, Bogdan S and Kowalski, Dariusz R and Rokicki, Mariusz A},
  booktitle={Proceedings IEEE INFOCOM},
  pages={1--5},
  year={2010},
  organization={IEEE}
}

@INPROCEEDINGS{aldawsari2019adversarial,
author={B. A. {Aldawsari} and B. S. {Chlebus} and D. R. {Kowalski}},
booktitle={Proceedings of the IEEE 18th International Symposium on Network Computing and Applications (NCA)},
title={Broadcasting on Adversarial Multiple Access Channels},
year={2019},
volume={},
number={},
pages={1-4},
}

@inproceedings{kowalski2005on,
author = {Kowalski, Dariusz R.},
title = {On selection problem in radio networks},
year = {2005},
isbn = {1581139942},
publisher = {Association for Computing Machinery},
address = {New York, NY, USA},
url = {https://doi.org/10.1145/1073814.1073843},
doi = {10.1145/1073814.1073843},
booktitle = {Proceedings of the Twenty-Fourth Annual ACM Symposium on Principles of Distributed Computing},
pages = {158–166},
numpages = {9},
location = {Las Vegas, NV, USA},
series = {PODC '05}
}

@inproceedings{jurdzinski2002energy,
 author = {Tomasz Jurdzi\'{n}ski and Miros\l{}aw Kuty\l{}owski and Jan Zatopia\'{n}ski},
 title = {Energy-Efficient Size Approximation of Radio Networks with No Collision Detection},
 booktitle = {Proceedings of the 8th Annual International Conference (COCOON)},
 year = {2002},
 pages = {279-289},
}

@inproceedings{bender2016contention,
 author = {Bender, Michael A. and Kopelowitz, Tsvi and Pettie, Seth and Young, Maxwell},
 title = {Contention Resolution with Log-logstar Channel Accesses},
 booktitle = {Proceedings of the 48th Annual ACM SIGACT Symposium on Theory of Computing},
 series = {STOC 2016},
 year = {2016},
 isbn = {978-1-4503-4132-5},
 location = {Cambridge, MA, USA},
 pages = {499-508},
 }

@inproceedings{bender2016how,
 author = {Bender, Michael A. and Fineman, Jeremy T. and Gilbert, Seth and Young, Maxwell},
 title = {{How to Scale Exponential Backoff: Constant Throughput, Polylog Access Attempts, and Robustness}},
 booktitle = {Proceedings of the Twenty-Seventh Annual ACM-SIAM Symposium on Discrete Algorithms},
 series = {SODA '16},
 year = {2016},
 pages = {636-654},
}

@inproceedings{awerbuch2008jamming,
    author = {Baruch Awerbuch and Andrea Richa and Christian Scheideler},
    title = {A Jamming-Resistant \textsc{MAC} Protocol for Single-Hop Wireless Networks},
    booktitle={Proceedings of the 27th ACM Symposium on Principles of Distributed Computing (PODC)},
    pages={45-54},
    year = {2008},
}

@INPROCEEDINGS{ogierman2014competitive, 
author={Adrian Ogierman and Andrea Richa and Christian Scheideler and Stefan Schmid and Jin Zhang}, 
booktitle={Proceedings of IEEE Conference on Computer Communications (INFOCOM)}, 
title={Competitive {MAC} under adversarial {SINR}}, 
year={2014}, 
pages={2751-2759}, 
}

@inproceedings{chen2021tight,
  author    = {Haimin Chen and
               Yonggang Jiang and
               Chaodong Zheng},
  title     = {Tight Trade-off in Contention Resolution without Collision Detection},
  booktitle = {Proceedings of the ACM Symposium on Principles of Distributed Computing},
  pages     = {139-149},
  year      = {2021},
  }

@inproceedings{anantharamu2011medium,
  title={Medium access control for adversarial channels with jamming},
  author={Anantharamu, Lakshmi and Chlebus, Bogdan S and Kowalski, Dariusz R and Rokicki, Mariusz A},
  booktitle={Proceedings of the International Colloquium on Structural Information and Communication Complexity},
  pages={89--100},
  year={2011},
  organization={Springer}
}

@inproceedings{bender2024fully,
  author       = {Michael A. Bender and
                  Jeremy T. Fineman and
                  Seth Gilbert and
                  John Kuszmaul and
                  Maxwell Young},
  title        = {Fully Energy-Efficient Randomized Backoff: Slow Feedback Loops Yield Fast Contention Resolution},
  booktitle    = {Proceedings of the 43rd {ACM} Symposium on Principles of Distributed Computing (PODC)},
  pages        = {231--242},
  publisher    = {{ACM}},
  year         = {2024},
}

@INPROCEEDINGS{caromi2019detection,
  author={Caromi, Raied and Souryal, Michael},
  booktitle={2019 Sensor Signal Processing for Defence Conference (SSPD)}, 
  title={Detection of Incumbent Radar in the 3.5 GHz CBRS Band using Support Vector Machines}, 
  year={2019},
  volume={},
  number={},
  pages={1-5},
  doi={10.1109/SSPD.2019.8751641}}

@article{bar2003windows,
  title={Windows scheduling problems for broadcast systems},
  author={Bar-Noy, Amotz and Ladner, Richard E},
  journal={SIAM Journal on Computing},
  volume={32},
  number={4},
  pages={1091--1113},
  year={2003},
  publisher={SIAM}
}

@article{bar2007windows,
  title={Windows scheduling as a restricted version of bin packing},
  author={Bar-Noy, Amotz and Ladner, Richard E and Tamir, Tami},
  journal={ACM Transactions on Algorithms (TALG)},
  volume={3},
  number={3},
  pages={28--es},
  year={2007},
  publisher={ACM New York, NY, USA}
}

@article{halper:station,
  title={Station Assignment with Reallocation},
  author={Halper, Austin and Mosteiro, Miguel A. and Rossikova, Yulia and  Wong, Prudence W. H. },
  journal={Algorithmica},
  volume={81},
  pages={1096--1125},
  year={2018},
}

@article{bohm2021packet,
  title={On packet scheduling with adversarial jamming and speedup},
  author={B{\"o}hm, Martin and Je{\.z}, {\L}ukasz and Sgall, Ji{\v{r}}{\'\i} and Vesel{\`y}, Pavel},
  journal={Annals of Operations Research},
  volume={298},
  number={1},
  pages={7--42},
  year={2021},
  publisher={Springer}
}

@inproceedings{chan2004line,
  title={On-line windows scheduling of temporary items},
  author={Chan, Wun-Tat and Wong, Prudence WH},
  booktitle={International Symposium on Algorithms and Computation},
  pages={259--270},
  year={2004},
  organization={Springer}
}

@article{farach2021dynamic,
  title={Dynamic Windows Scheduling with Reallocation},
  author={Farach-Colton, Mart{\'\i}n and Leal, Katia and Mosteiro, Miguel A and Caro, Christopher Thraves},
  journal={Journal of Experimental Algorithmics (JEA)},
  volume={26},
  pages={1--19},
  year={2021},
  publisher={ACM New York, NY, USA}
}

@inproceedings{agrawal2020contention,
    author = {Agrawal, Kunal and Bender, Michael A. and Fineman, Jeremy T. and Gilbert, Seth and Young, Maxwell},
    title = {Contention Resolution with Message Deadlines},
    year = {2020},
    isbn = {9781450369350},
    publisher = {Association for Computing Machinery},
    address = {New York, NY, USA},
    url = {https://doi.org/10.1145/3350755.3400239},
    doi = {10.1145/3350755.3400239},
    booktitle = {Proceedings of the 32nd ACM Symposium on Parallelism in Algorithms and Architectures},
    pages = {23–35},
    numpages = {13},
    location = {Virtual Event, USA},
    series = {SPAA '20}
}

@article{borodin2001adversarial,
  title={Adversarial queuing theory},
  author={Borodin, Allan and Kleinberg, Jon and Raghavan, Prabhakar and Sudan, Madhu and Williamson, David P},
  journal={Journal of the ACM (JACM)},
  volume={48},
  number={1},
  pages={13--38},
  year={2001},
  publisher={ACM New York, NY, USA}
}

@article{chlebus2009maximum,
  title={Maximum throughput of multiple access channels in adversarial environments},
  author={Chlebus, Bogdan S and Kowalski, Dariusz R and Rokicki, Mariusz A},
  journal={Distributed Computing},
  volume={22},
  number={2},
  pages={93--116},
  year={2009},
  publisher={Springer}
}

@ARTICLE{komlos1985asymptotically,
  author={Komlos, J. and Greenberg, A.},
  journal={IEEE Transactions on Information Theory}, 
  title={An asymptotically fast nonadaptive algorithm for conflict resolution in multiple-access channels}, 
  year={1985},
  volume={31},
  number={2},
  pages={302-306},
  doi={10.1109/TIT.1985.1057020}}

@article{banicescu2024survey,
author = {Banicescu, Ioana and Chakraborty, Trisha and Gilbert, Seth and Young, Maxwell},
title = {A Survey on Adversarial Contention Resolution},
year = {2025},
issue_date = {November 2025},
publisher = {Association for Computing Machinery},
address = {New York, NY, USA},
volume = {57},
number = {11},
journal = {ACM Computing Surveys},
month = jun,
articleno = {281},
numpages = {37},
}

@article{anantharamu2019packet,
  title={Packet latency of deterministic broadcasting in adversarial multiple access channels},
  author={Anantharamu, Lakshmi and Chlebus, Bogdan S and Kowalski, Dariusz R and Rokicki, Mariusz A},
  journal={Journal of Computer and System Sciences},
  volume={99},
  pages={27-52},
  year={2019},
  publisher={Elsevier}
}

@article{bender2025jamming,
  author       = {Michael A. Bender and
                  Jeremy T. Fineman and
                  Seth Gilbert and
                  John Kuszmaul and
                  Maxwell Young},
  title        = {Jamming-Resistant Backoff with Polylogarithmic Sending and Listening
                  Cost},
  journal      = {{SIAM} Journal on Computing},
  volume       = {54},
  number       = {5},
  pages        = {1335--1385},
  year         = {2025},
}

@article{roughgarden2020resource,
  author       = {Tim Roughgarden},
  title        = {Resource Augmentation},
  journal      = {CoRR},
  volume       = {abs/2007.13234},
  year         = {2020},
  url          = {https://arxiv.org/abs/2007.13234},
  eprinttype    = {arXiv},
  eprint       = {2007.13234},
  bibsource    = {dblp computer science bibliography, https://dblp.org}
}

@ARTICLE{bhattarai2016overview,
  author={Bhattarai, Sudeep and Park, Jung-Min Jerry and Gao, Bo and Bian, Kaigui and Lehr, William},
  journal={IEEE Transactions on Cognitive Communications and Networking}, 
  title={An Overview of Dynamic Spectrum Sharing: Ongoing Initiatives, Challenges, and a Roadmap for Future Research}, 
  year={2016},
  volume={2},
  number={2},
  pages={110-128},
  doi={10.1109/TCCN.2016.2592921}}

@ARTICLE{peha1998spectrum,
  author={Peha, Jon M.},
  journal={IEEE Communications Surveys}, 
  title={Spectrum management policy options}, 
  year={1998},
  volume={1},
  number={1},
  pages={2-8},
  doi={10.1109/COMST.1998.5340403}}

@article{wen1998preemptive,
    title = {Preemptive on-line scheduling for two uniform processors},
    journal = {Operations Research Letters},
    volume = {23},
    number = {3},
    pages = {113-116},
    year = {1998},
    issn = {0167-6377},
    doi = {https://doi.org/10.1016/S0167-6377(98)00032-7},
    url = {https://www.sciencedirect.com/science/article/pii/S0167637798000327},
    author = {Jianjun Wen and Donglei Du},
}

@article{dosa2018tight,
  author    = {Gy{\"o}rgy D{\'o}sa and Armin F{\"u}genschuh and Zsolt Tan and Zsolt Tuza and Krzysztof W{\k{e}}sek},
  title     = {Tight Upper Bounds for Semi-Online Scheduling on Two Uniform Machines with Known Optimum},
  journal   = {Central European Journal of Operations Research},
  volume    = {26},
  number    = {1},
  pages     = {161--180},
  year      = {2018},
  publisher = {Springer},
  doi       = {10.1007/s10100-017-0481-z}
}

@article{zheng2024semi,
	author = {{Zheng, Feifeng} and {Chen, Yuhong} and {Liu, Ming}},
	title = {Semi-online scheduling on two uniform parallel machines with initial lookahead},
	DOI= {10.1051/ro/2024042},
	url= {https://doi.org/10.1051/ro/2024042},
	journal = {RAIRO-Oper. Res.},
	year = {2024},
	volume = {58},
	number = {3},
	pages = {2621-2630},
}

@article{epstein2014preemptive,
    title = "Preemptive online scheduling with rejection of unit jobs on two uniformly related machines",
    author = "Leah Epstein and Hanan Zebedat-Haider",
    year = "2014",
    month = "feb",
    doi = "10.1007/s10951-013-0326-0",
    language = "English",
    volume = "17",
    pages = "87--93",
    journal = "Journal of Scheduling",
    issn = "1094-6136",
    publisher = "Springer",
    number = "1",
}

@article{dosa2011online,
    title = {Online scheduling with rearrangement on two related machines},
    journal = {Theoretical Computer Science},
    volume = {412},
    number = {8},
    pages = {642-653},
    year = {2011},
    issn = {0304-3975},
    doi = {https://doi.org/10.1016/j.tcs.2010.10.019},
    url = {https://www.sciencedirect.com/science/article/pii/S0304397510005815},
    author = {György Dósa and Yuxin Wang and Xin Han and He Guo},
}

@article{maack2023online,
    title = {Online load balancing on uniform machines with limited migration},
    journal = {Operations Research Letters},
    volume = {51},
    number = {3},
    pages = {220-225},
    year = {2023},
    issn = {0167-6377},
    doi = {https://doi.org/10.1016/j.orl.2023.02.013},
    url = {https://www.sciencedirect.com/science/article/pii/S0167637723000408},
    author = {Marten Maack},
}

@techreport{ida2015spectrum_sharing,
  title        = {Federal Spectrum Sharing: An Assessment of the Technology and Opportunities},
  author       = {{Institute for Defense Analyses}},
  institution  = {Institute for Defense Analyses (IDA)},
  year         = {2015},
  number       = {P-5186},
  url          = {https://www.ida.org/~/media/Corporate/Files/Publications/STPIPubs/2015/p5186final.pdf},
  note         = {Accessed: 2026-02-13}
}

@techreport{calabrese2021useitorshare,
  author       = {Calabrese, Michael A.},
  title        = {Use it or Share it: A New Default Policy for Spectrum Management},
  institution  = {Open Technology Institute at New America},
  year         = {2021},
  month        = {February},
  note         = {Presented at the 48th Research Conference on Communication, Information and Internet Policy (TPRC48)},
  url          = {https://ssrn.com/abstract=3762098},
}

@misc{fcc_sas_35ghz,
  author = {{Federal Communications Commission}},
  title  = {3.5 {GHz} Band Overview: Spectrum Access System ({SAS})},
  year   = {2024},
  note   = {Available on the FCC website.}
}

\clearpage
\appendix

\section{Supporting Analysis}\label{sec:upper-bound-analysis}

This appendix provides execution details and full proofs supporting the main results.

\subsection{Execution Details}\label{sec:execution-details}

\noindent{\bf \tstb Execution Windows.}
We describe how \mainAlgorithmAcrynom uses \TSTB (\tstb)  \cite{bender2016contention}. For an integer parameter $z$, a \tstb execution consists of a predetermined sequence of contention windows containing a total of $L(z)=O(z)$ slots. The first window contains $2z$ slots. On the fast channel, each such slot corresponds to one fast-channel mini-slot.  Each active packet independently selects one  slot uniformly at random in each window, and successful packets depart. A fresh \tstb execution with $z$ packets leaves at least one packet unfinished with probability $O(z^{-d})$, for some fixed $d\geq2$. If necessary, this guarantee can be obtained through a constant number of repetitions without changing the execution length $L(z)=O(z)$.

A \tstb execution can be paused and later resumed from where it stopped, without restarting its sequence of contention windows. Each \tstb slot corresponds to one slow-channel slot when executed on the slow channel, or one fast-channel mini-slot when executed on a logical fast subchannel.\medskip

\noindent{\bf Reporting Evictions.}
Every fast round lasts for its scheduled $D(x)$ slots, even if all selected packets succeed earlier. Packets evicted before the scheduled end return to the slow channel and wait. At the end of the scheduled execution, any packets still on the fast channel also return to the slow channel.

The next slow-channel slot is left idle to allow all remaining packets to return. During the following slot, packets that were evicted before the scheduled end and have not yet succeeded transmit to report the eviction. All other packets listen. An empty report certifies the round, while a nonempty report aborts it. If exactly one packet reports, its transmission succeeds, but $W$ is not incremented. Thus, every remaining packet learns whether the round was certified or aborted, even if the fast channel is no longer available.

\medskip
\noindent{\bf Cleanup.}
When either cutoff is reached, \mainAlgorithmAcrynom permanently stops using the fast channel. A fast round already in progress finishes before cleanup begins, whereas an ongoing call to \estimateFunction\ is stopped if the time cutoff is reached. If the initial estimation finishes after the time cutoff, \mainAlgorithmAcrynom enters cleanup immediately.

During cleanup, the remaining packets execute \tstb on the slow channel with successively doubled size parameters $s_0,2s_0,4s_0,\ldots$, until all packets succeed. The initial parameter is $s_0=\max\{v,s_{\min}\}$, where $s_{\min}$ is a sufficiently large fixed constant. Here $v=\widehat{n}_{\mathrm{cur}}$ when cleanup is triggered by the population cutoff, and $v=\widehat{n}$ when triggered by the time cutoff. Lemma~\ref{lem:cleanup} establishes that, starting with $r$ active packets, the expected cleanup time is $O(r+v)$.

\medskip
\noindent{\bf Minimum Fast-Channel Availability.}
The analysis requires each interval of fast-channel availability to contain sufficiently many slots to accommodate consecutive rounds of intended per-subchannel size $\rho$, including their reporting slots. Specifically, it suffices to require $C_{\mathrm{avail}}\geq 2D(\rho)+8$. 
Since $D(\rho)=O(1)$, this requires only a fixed constant number of slots. This assumption is used in Lemma~\ref{lem:baseline} to bound the number of aborted rounds.

\subsection{Fresh-Window Estimates}\label{sec:helper-lemmas}

We begin with two properties of fast-channel batches: a bound on the expected number of packets that remain unsuccessful after an uninterrupted \tstb execution, and a high-probability guarantee on the number of packets that succeed.

\begin{lemma}
\label{lem:return}\label{lem:non-evicted-batch-cost}
Consider a fast round that begins with $r\leq\widehat{n}_{\mathrm{cur}}$ active packets. For any one logical fast subchannel, let $Z$ be the number of packets that would remain unsuccessful after its full scheduled \tstb execution, assuming no eviction occurs. Then
$$
E[Z]=O(1),
$$
where the expectation is over the random choices made during the round.
\end{lemma}

\begin{proof}
Let $Y$ be the number of packets assigned to the subchannel. Since each active packet independently selects it with probability $x/\widehat{n}_{\mathrm{cur}}$, we have:
$$
E[Y]=\frac{rx}{\widehat{n}_{\mathrm{cur}}}\leq x.
$$

First, suppose $Y\leq2x$. Compare the \tstb execution with one containing exactly $2x$ packets by adding imaginary packets. Each real packet makes the same random choices in both executions. Adding packets cannot help a real packet succeed earlier, so the number of unfinished real packets is at most the number left unfinished in the larger execution.

Since \tstb runs with parameter $2x$, its completion guarantee implies that the larger execution leaves any packet unfinished with probability $O(x^{-d})$, where $d\geq2$. Consequently, the contribution to $E[Z]$ from $Y\leq2x$ is at most:
$$
2x\cdot O(x^{-d})=O(1).
$$

Now suppose $Y>2x$. By a standard Chernoff bound:
$$
\Pr(Y>2x)\leq\exp(-\Omega(x)).
$$
Since $Y$ is binomial, $\mathrm{Var}(Y)\leq E[Y]\leq x$. By the definition of variance:
$$
\mathrm{Var}(Y)=E[(Y-E[Y])^2]=E[Y^2]-(E[Y])^2.
$$
Rearranging gives:
$$
E[Y^2]=\mathrm{Var}(Y)+(E[Y])^2\leq x+x^2.
$$

We now use Cauchy--Schwarz to bound the contribution to $E[Y]$ from the event $Y>2x$. Applied to the conditional distribution given this event, Cauchy--Schwarz gives:
$$
(E[Y\mid Y>2x])^2\leq E[Y^2\mid Y>2x].
$$
Multiplying by $\Pr(Y>2x)^2$ and taking square roots, we obtain:
$$
E[Y\mid Y>2x]\Pr(Y>2x)
\leq
\sqrt{E[Y^2\mid Y>2x]\Pr(Y>2x)^2}.
$$
Since the event $Y>2x$ contributes at most $E[Y^2]$ to the second moment, we also have:
$$
E[Y^2\mid Y>2x]\Pr(Y>2x)\leq E[Y^2].
$$
Combining these inequalities gives:
$$
E[Y\mid Y>2x]\Pr(Y>2x)
\leq\sqrt{E[Y^2]\Pr(Y>2x)}
\leq\sqrt{(x+x^2)\exp(-\Omega(x))}
=O(1).
$$
Since $Z\leq Y$, this also bounds the contribution to $E[Z]$ from $Y>2x$.

Finally, combining the two cases gives $E[Z]=O(1)$.
\end{proof}

\begin{lemma}
\label{lem:witness}\label{lem:exp-number-revised}
Let $K=64$. Consider a fast-channel round that begins with $r$ active packets satisfying:
$$
\frac{\widehat{n}_{\mathrm{cur}}}{2K}
\leq r\leq\widehat{n}_{\mathrm{cur}}.
$$
With probability at least $1-O(\lambda e^{-\Omega(x)})$, each logical fast subchannel receives between $x/(4K)$ and $2x$ packets. Moreover, if the scheduled \tstb execution on each subchannel runs to completion without interruption, at least $\alpha x$ of its packets succeed, where $\alpha>0$ is a fixed constant independent of $\lambda$. 
\end{lemma}
\begin{proof}
Consider one logical fast subchannel, and let $Y$ be the number of packets assigned to it. Each of the $r$ active packets independently chooses this subchannel with probability $x/\widehat{n}_{\mathrm{cur}}$. Consequently, $Y$ is a binomial random variable with expectation:
$$
E[Y]=\frac{rx}{\widehat{n}_{\mathrm{cur}}}.
$$
By the assumption on $r$, we have:
$$
\frac{\widehat{n}_{\mathrm{cur}}}{2K}
\leq r\leq\widehat{n}_{\mathrm{cur}}.
$$
Multiplying by $x/\widehat{n}_{\mathrm{cur}}$ gives:
$$
\frac{x}{2K}\leq E[Y]\leq x.
$$
We first show that the number of packets assigned to the subchannel is between $x/(4K)$ and $2x$ with high probability. Since $E[Y]\geq x/(2K)$, a standard Chernoff lower-tail bound gives:
$$
\Pr\left(Y<\frac{x}{4K}\right)
\leq\exp(-\Omega(x)).
$$
Similarly, since $E[Y]\leq x$, a Chernoff upper-tail bound gives:
$$
\Pr(Y>2x)\leq\exp(-\Omega(x)).
$$
Thus, with probability at least $1-\exp(-\Omega(x))$, the number of packets assigned to the subchannel satisfies:
$$
\frac{x}{4K}\leq Y\leq2x.
$$
We now show that a constant fraction of these packets succeed during an uninterrupted \tstb execution. Suppose exactly $y$ packets have been assigned to the subchannel, where:
$$
\frac{x}{4K}\leq y\leq2x.
$$
The \tstb execution uses parameter $2x$, so its first contention window contains $4x$ mini-slots. Each of the $y$ packets independently chooses one of these mini-slots uniformly at random.

Let $Z$ be the number of packets that choose a mini-slot not chosen by any other packet. Each such packet transmits successfully during the first window. For any particular packet, the probability that none of the other $y-1$ packets chooses its mini-slot is:
$$
\left(1-\frac{1}{4x}\right)^{y-1}.
$$
By linearity of expectation:
$$
E[Z]=y\left(1-\frac{1}{4x}\right)^{y-1}.
$$
For any particular packet, a union bound implies that the probability of at least one other packet choosing the same mini-slot is at most $(y-1)/(4x)$. Since $y\leq2x$, the probability that no other packet chooses its mini-slot is therefore at least:
$$
1-\frac{y-1}{4x}\geq\frac12,
$$
and consequently $E[Z]\geq y/2$. 

We next show that $Z$ is unlikely to be much smaller than its expectation. We use McDiarmid's bounded-differences inequality, which applies to functions of independent random variables. The $y$ independently chosen mini-slots determine $Z$. Suppose we change the mini-slot selected by one packet while keeping all other packets' choices fixed. This can affect the number of successful transmissions in at most two mini-slots: the one originally selected and the new one. In each, the number of successes can change by at most one. Thus, $Z$ can change by at most two. Thus, the bounded-differences inequality gives, for any $u>0$:
$$
\Pr(Z\leq E[Z]-u)
\leq\exp\left(-\frac{2u^2}{4y}\right).
$$
Since $E[Z]\geq y/2$, choosing $u=y/4$ yields:
$$
\Pr\left(Z<\frac y4\right)
\leq\exp\left(-\frac{y}{32}\right).
$$
Because $y\geq x/(4K)$, this probability is $\exp(-\Omega(x))$. Hence, except with this probability, at least:
$$
\frac y4\geq\frac{x}{16K}
$$
packets succeed during the first \tstb window.

Combining this result with the Chernoff bounds, we conclude that a particular logical fast subchannel receives between $x/(4K)$ and $2x$ packets and completes at least $\alpha x$ packets during an uninterrupted execution, except with probability $\exp(-\Omega(x))$, for some fixed constant $\alpha>0$ independent of $\lambda$.

Finally, \mainAlgorithmAcrynom uses $\lambda$ logical fast subchannels. Applying a union bound over these subchannels, the probability that any subchannel fails either guarantee is at most $O(\lambda e^{-\Omega(x)})$, which completes the argument.
\end{proof}

\noindent{\bf Consequences of Lemma~\ref{lem:exp-number-revised} for Fast Rounds.}
Let $G$ denote the event that all population estimates satisfy their accuracy guarantees and that the packet-assignment and progress guarantees of Lemma~\ref{lem:exp-number-revised} hold for every fast round to which that lemma applies. Later on, Lemmas~\ref{lem:invariant} and~\ref{lem:good} establish that, on $G$, the lemma applies to every fast round and that $G$ occurs with high probability.

The preceding lemma therefore implies that, on event $G$, each certified round completes at least $\alpha\lambda x$ packets. This follows because either the round was not interrupted by an eviction, or an eviction occurred after all selected packets had already succeeded. Otherwise, an unfinished packet would have reported the eviction, causing the round to be aborted. On $G$, the lemma also guarantees that at most $2\lambda x$ packets are selected for any round, whether certified or aborted.

\subsection{Charging Aborted Rounds}\label{sec:aborts}

Recall that $D(\rho)=O(1)$ and that $C_{\mathrm{avail}}$ was chosen to satisfy $C_{\mathrm{avail}}\geq2D(\rho)+8$ for every integer $\rho\geq2$. A necessary portion of our argument is that the minimum duration of fast-channel availability guarantees that, after an aborted round, \mainAlgorithmAcrynom has enough time to complete its next round with $x=\rho$ without another eviction, unless it invokes the population estimator in between. 

\begin{lemma}\label{lem:baseline}
Consider two fast rounds, each with an intended batch size of $x=\rho$ packets per logical subchannel. If no certified round or invocation of \estimateFunction\ occurs between them, then the two rounds cannot both be aborted.
\end{lemma}
\begin{proof}
Suppose two fast rounds with $x=\rho$ occur without a certified round or an invocation of \estimateFunction\ between them. We show that if the first round is aborted, the second cannot be aborted.

The key observation is that the first round may begin near the end of an interval of fast-channel availability, leaving insufficient time to finish before an eviction. However, after this eviction, the next round with $x=\rho$ begins sufficiently early in a new interval of availability to finish without interruption.

Let the first round begin at slow-slot boundary $t$. Since it is aborted, an eviction must have occurred before its scheduled $D(\rho)$ slots of \tstb execution finished. Nevertheless, \mainAlgorithmAcrynom waits until the scheduled execution ends and then uses two additional slots to determine whether an eviction occurred. Thus, the first round ends at time $t+D(\rho)+2$, and \mainAlgorithmAcrynom cannot begin another fast round before then.

We consider two cases, depending on when the fast channel becomes available again after the eviction that interrupted the first round.\smallskip

\noindent\emph{Case 1: After the eviction, the fast channel becomes available again before the first round ends.}
The new interval of fast-channel availability begins no earlier than $t$, so at most $D(\rho)+2$ slow slots have elapsed between the beginning of that interval and the end of the first round. Since every interval of fast-channel availability contains at least $C_{\mathrm{avail}}\geq2D(\rho)+8$ complete slow slots, enough time remains for \mainAlgorithmAcrynom to execute another $D(\rho)$ slots of \tstb without interruption. Thus, if \mainAlgorithmAcrynom begins the second round, it completes its scheduled \tstb execution before another eviction can occur.\smallskip

\noindent\emph{Case 2: After the eviction, the fast channel becomes available again at or after the first round ends.}
\mainAlgorithmAcrynom observes that the fast channel is available at the next slow-slot boundary. Because no population estimate or certified round intervenes, \mainAlgorithmAcrynom begins its next fast round at that boundary, provided it has not entered cleanup. At most one slow slot is lost between the beginning of the new interval of availability and the start of this round. The minimum-availability condition therefore guarantees that the fast channel remains available throughout its scheduled $D(\rho)$ slots of \tstb execution.\smallskip

In either case, the second round completes its scheduled \tstb execution without eviction and therefore cannot be aborted. If the fast channel never becomes available again, or \mainAlgorithmAcrynom enters cleanup, no second round occurs. This completes the argument.
\end{proof}

Recall that an \emph{estimation epoch} begins when \mainAlgorithmAcrynom obtains a new estimate of the number of active packets and ends when \mainAlgorithmAcrynom next invokes \estimateFunction\ or enters permanent slow-channel cleanup. At the beginning of each epoch, $W$ is reset to zero.

In the following lemma, $F$ denotes an upper bound on the intended number of packets per subchannel during the fast rounds being considered. We will specify a suitable value of $F$ later in the analysis.

\begin{lemma}
\label{lem:charge}
Consider \mainAlgorithmAcrynom's execution up to any time $t$ before cleanup. Suppose every fast round has intended per-subchannel batch size $x=\rho+b\leq F$. Let $h$ be the number of calls to \estimateFunction\ that have started after the initial estimate and by time $t$, including any call still in progress.

Let $V$ and $I$ be the sums of $x$ over certified and aborted rounds, respectively, whose reporting slots have finished by time $t$. Then:
$$
I\leq4V+\rho(h+1).
$$
Within any single estimation epoch $j$, the corresponding sums satisfy:
$$
I_j\leq4V_j+4F.
$$
\end{lemma}

\begin{proof}
We first establish the bound over the entire execution and we consider three cases.\medskip

\noindent{\it Case 1.} Consider an aborted round with intended per-subchannel batch size $x=\rho+t$, where $t\geq1$. Since an abort resets $b$ to zero, \mainAlgorithmAcrynom must have previously completed $t$ certified rounds with intended per-subchannel batch sizes $\rho,\rho+1,\ldots,\rho+t-1$ since the preceding abort. The sum of these sizes is at least $t\rho$. Moreover, since $\rho\geq1$ and $t\geq1$, we have:
$$
\rho+t\leq2t\rho
\leq2\sum_{i=0}^{t-1}(\rho+i).
$$
Thus, the intended size of the aborted round is at most twice the sum of the intended sizes of these preceding certified rounds. Since the certified rounds used in this argument are different for each aborted round, the total intended size of all aborted rounds with $x>\rho$ is at most $2V$.\medskip

\noindent{\it Case 2.} We must also account for aborted rounds with $x=\rho$. Consider such a round that follows an earlier aborted round with $x>\rho$, without any certified round between them. Its intended size $\rho$ is at most that of the earlier aborted round, so the total contribution from these rounds is bounded by the contribution already established above in Case 1.\medskip

\noindent{\it Case 3.} The remaining possibility is an aborted round with $x=\rho$ that either occurs first or follows another aborted round with $x=\rho$. By Lemma~\ref{lem:baseline}, the latter situation requires an intervening call to \estimateFunction. Hence, there can be at most one such additional aborted round for each call to \estimateFunction, plus one at the beginning of the execution. Their combined intended size is at most $\rho(h+1)$.\medskip

\noindent Combining these bounds gives:
$$
I\leq4V+\rho(h+1).
$$

We next establish the bound within a single estimation epoch $j$. The argument is similar, except that the first aborted round with $x>\rho$ may have been preceded by certified rounds belonging to an earlier epoch. We therefore account for this round separately, using the assumption that its intended per-subchannel batch size is at most $F$.

For every subsequent aborted round with $x>\rho$, the preceding certified rounds needed to reach its batch size belong to the current epoch. As before, their intended sizes account for at most twice their sum in aborted rounds. Consequently, the sum of the intended sizes of aborted rounds with $x>\rho$ during epoch $j$ is at most $2V_j+F$.

Finally, each aborted round with $x=\rho$ can be accounted for by an earlier aborted round with $x>\rho$, with at most one exception per epoch by Lemma~\ref{lem:baseline}. Thus, the total intended size of aborted rounds with $x=\rho$ is at most $2V_j+F+\rho$.

Combining these contributions, and using $\rho\leq F$, gives:
$$
I_j\leq4V_j+2F+\rho
\leq4V_j+4F,
$$
which completes the proof.
\end{proof}

\subsection{Accuracy and Measurement Frequency}\label{sec:invariant}

We now establish that \mainAlgorithmAcrynom maintains an accurate estimate of the number of active packets throughout its execution, and that only $O(\log n)$ population estimates are needed. We also show that the intended fast-batch sizes remain sufficiently small that the limit imposed by the current population estimate does not affect the increasing sequence of batch sizes.

Let $G_0$ denote the event that the initial call to \estimateFunction\ returns an estimate satisfying $n\leq\widehat{n}\leq32n$ and finishes in $O(\log^2 n)$ slots. We establish that this event occurs with high probability in Appendix~\ref{sec:estimate-n}.

For the analysis, let $F$ be a sufficiently large constant multiple of $\rho+\sqrt{n/\lambda}$, where the constant is independent of $\lambda$. Thus:
$$
F=C_F\left(\rho+\sqrt{n/\lambda}\right)=O(\sqrt n).
$$
where $C_F>0$ is a sufficiently large fixed constant independent of $\lambda$.  We will show that $F$ bounds the intended number of packets per subchannel in every fast round. This bound is used only in the analysis and does not affect the algorithm.

As in Lemma~\ref{lem:exp-number-revised}, we use $K=64$ as a fixed constant in our bounds on the accuracy of the population estimates. Recall that $\delta>0$ determines when \mainAlgorithmAcrynom requests a new estimate, $W\geq\delta\widehat{n}_{\mathrm{cur}}$, and the constant $c_{\mathrm{pop}}>0$ determines the population cutoff $\tau_{\mathrm{pop}}=\lceil c_{\mathrm{pop}}\widehat{n}/\rho\rceil$. To make the requirements on these constants clear, we choose $\delta$ sufficiently small and $c_{\mathrm{pop}}$ sufficiently large so that:
\begin{equation}\label{eq:invariant-constants}
10\delta\leq\frac{1}{8K},\qquad
20\lambda F+20\leq\frac{\tau_{\mathrm{pop}}}{8K},
\qquad
\tau_{\mathrm{pop}}\geq2\lambda\rho.
\end{equation}
These conditions ensure that the number of packets completing between population estimates remains small enough for the current estimate to stay accurate. On $G_0$, we also have $\tau_{\mathrm{est}}(\widehat{n})\leq\tau_{\mathrm{pop}}/(4K)$ for sufficiently large $n$. We justify these choices of constants later on in this analysis.

Finally, let $G$ denote the event that the initial estimate satisfies the guarantees above, every subsequent completed call to \estimateFunction\ satisfies its accuracy guarantee, and the packet-assignment and \tstb progress guarantees of Lemma~\ref{lem:exp-number-revised} hold whenever a fast round begins with $r$ active packets satisfying:
$$
\frac{\widehat{n}_{\mathrm{cur}}}{2K}
\leq r\leq\widehat{n}_{\mathrm{cur}}.
$$
Here, the \tstb progress guarantee concerns the packets that would succeed if the scheduled execution were not interrupted by an eviction. 

The next lemma establishes that, on event $G$, \mainAlgorithmAcrynom maintains a sufficiently accurate estimate of the number of active packets throughout its execution before cleanup. It also shows that the intended number of packets per subchannel remains below $F$ and follows $x=\rho+b$, without being restricted by the current population estimate. Together, these properties allow us to apply the fast-channel progress guarantees of Lemma~\ref{lem:exp-number-revised} throughout \mainAlgorithmAcrynom's execution.

\begin{lemma}
\label{lem:invariant}
On event $G$, let $N_j$ and $\widehat{n}_{\mathrm{cur},j}$ denote the actual number of active packets and \mainAlgorithmAcrynom's estimate of that number, respectively, at the beginning of estimation epoch $j$. Then:
$$
N_j\leq\widehat{n}_{\mathrm{cur},j}\leq K N_j.
$$
Moreover, whenever \mainAlgorithmAcrynom begins a fast round with $r$ active packets, we have:
$$
\frac{\widehat{n}_{\mathrm{cur}}}{2K}
\leq r\leq\widehat{n}_{\mathrm{cur}}.
$$
Finally, the intended number  of packets per subchannel satisfies $x=\rho+b<F$, where $F$ is the upper bound defined above. Consequently, whenever \mainAlgorithmAcrynom begins a fast round, the intended number of packets per subchannel is $x=\rho+b$: the upper limit $\lfloor\widehat{n}_{\mathrm{cur}}/\lambda\rfloor$ in the batch-size formula never restricts $x$ before cleanup.
\end{lemma}
\begin{proof}
We prove the claims simultaneously by induction over successive fast rounds and population-estimate updates. The base case follows from the initial population-estimation guarantee described in Section~\ref{sec:primitive}, whose proof appears in Appendix~\ref{sec:estimate-n}. For the induction, we assume the claims hold for all preceding rounds and estimates. In particular, the estimates were sufficiently accurate for Lemma~\ref{lem:exp-number-revised} to apply to those earlier rounds. We show that their progress and the bounds on their sizes preserve the required guarantees when \mainAlgorithmAcrynom begins another fast round or obtains a new population estimate.

On event $G_0$, the initial estimate satisfies $n\leq\widehat{n}\leq32n$. Since the initial call to \estimateFunction\ takes $O(\log^2 n)$ slots, at least $n/2$ packets remain active when it finishes, for sufficiently large $n$. Thus, if $N_0$ is the number of active packets at the beginning of the first estimation epoch, then:
$$
N_0\leq\widehat{n}\leq K N_0,
$$
where $K=64$.

\medskip
\noindent{\it Bounding fast-channel batch sizes.}
Suppose the claims have held for all preceding fast rounds and population estimates. If the current batch index is $b$, then \mainAlgorithmAcrynom has certified $b$ rounds since the last reported abort. In these rounds, the intended number of packets per subchannel increases through the sequence $\rho,\rho+1,\ldots,\rho+b-1$. On event $G$, Lemma~\ref{lem:exp-number-revised} guarantees that these rounds complete at least:
$$
\alpha\lambda\sum_{i=0}^{b-1}(\rho+i)
\geq\frac{\alpha\lambda b^2}{2}
$$
distinct packets. Since there are only $n$ packets, we obtain:
$$
b\leq\sqrt{\frac{2n}{\alpha\lambda}}.
$$
Thus $\rho+b<F$, by our choice of $F$. Furthermore, before cleanup, $\widehat{n}_{\mathrm{cur}}>\tau_{\mathrm{pop}}$, and our choice of constants in~\eqref{eq:invariant-constants} guarantees:
$$
\left\lfloor\frac{\widehat{n}_{\mathrm{cur}}}{\lambda}\right\rfloor>F.
$$
Consequently, whenever \mainAlgorithmAcrynom begins another fast round, its intended number of packets per subchannel is $x=\rho+b<F$.

\medskip
\noindent{\it Maintaining accurate population estimates.}
Consider estimation epoch $j$, beginning with $N_j$ active packets and an estimate satisfying $N_j\leq\widehat{n}_{\mathrm{cur},j}\leq K N_j$.

Let $C_s$ count slow-channel \tstb successes during this epoch, and let $V_j$ and $I_j$ be the sums of the intended per-subchannel batch sizes over certified and aborted rounds, respectively. Recall that $W=C_s+\lambda V_j$.

By the induction hypothesis, preceding fast rounds began with sufficiently accurate population estimates. Thus, on event $G$, Lemma~\ref{lem:exp-number-revised} guarantees that at most $2\lambda x$ packets participated in each of those rounds. Since $V_j+I_j$ sums their intended per-subchannel batch sizes, at most $2\lambda(V_j+I_j)$ packets could have succeeded during these rounds.

We must also account for a round that may still be underway, whose successful packets are not yet included in these sums. Since its intended per-subchannel batch size is at most $F$, it can contribute at most another $2\lambda F$ successful packets on $G$. Including the $C_s$ slow-channel \tstb successes, and using $I_j\leq4V_j+4F$ from Lemma~\ref{lem:charge} together with $W=C_s+\lambda V_j$, we obtain:
\begin{equation}\label{eq:departure-bound}
C_s+2\lambda(V_j+I_j)+2\lambda F
\leq10W+10\lambda F.
\end{equation}

\mainAlgorithmAcrynom requests a new estimate when $W$ reaches $\delta\widehat{n}_{\mathrm{cur},j}$. Because each increase in $W$ is at most $\lambda F+1$, the preceding bound implies that, before the next estimate begins, at most:
$$
10\delta\widehat{n}_{\mathrm{cur},j}+20\lambda F+20
\leq\frac{\widehat{n}_{\mathrm{cur},j}}{4K}
$$
packets have succeeded, where the final inequality follows from~\eqref{eq:invariant-constants}.

Since at least $\widehat{n}_{\mathrm{cur},j}/K$ packets were active at the beginning of the epoch, the number $r$ of active packets before the next estimate satisfies:
$$
\frac{\widehat{n}_{\mathrm{cur},j}}{2K}
\leq r\leq\widehat{n}_{\mathrm{cur},j}.
$$
Thus the estimate remains sufficiently accurate for Lemma~\ref{lem:exp-number-revised} to apply whenever \mainAlgorithmAcrynom begins another fast round.

Finally, when \mainAlgorithmAcrynom requests a new estimate, at least:
$$
\frac{\widehat{n}_{\mathrm{cur},j}}{2K}
\geq\frac{\tau_{\mathrm{pop}}}{2K}
\geq2\tau_{\mathrm{est}}(\widehat{n})
$$
packets remain active. Lemma~\ref{lem:estimate} therefore guarantees that, unless all remaining packets succeed during estimation, the new estimate is within a factor of $16$ of the actual number of active packets when the estimator finishes. Since $K=64$, the required accuracy guarantee holds at the beginning of the next epoch. If the time cutoff interrupts the estimator, \mainAlgorithmAcrynom enters cleanup instead. This completes the induction argument and establishes all three claims.
\end{proof}

\begin{lemma}
\label{lem:epochs}
On event $G$, \mainAlgorithmAcrynom invokes \estimateFunction\ at most $O(\log n)$ times after the initial estimate. Moreover, if $\widehat{n}_{\mathrm{cur},j}$ denotes the estimate at the beginning of estimation epoch $j$, then:
$$
\sum_j\widehat{n}_{\mathrm{cur},j}=O(n).
$$
\end{lemma}
\begin{proof}
Consider an estimation epoch $j$ that ends when \mainAlgorithmAcrynom invokes \estimateFunction. At that point, the counter satisfies:
$$
W=C_s+\lambda V_j\geq\delta\widehat{n}_{\mathrm{cur},j},
$$
where $C_s$ counts successful slow-channel \tstb transmissions and $V_j$ is the sum of the intended per-subchannel batch sizes over certified rounds during the epoch.

On event $G$, each certified round completes at least an $\alpha$ fraction of the number of packets credited to $W$. Therefore, the number of packets that have succeeded during epoch $j$ is at least:
$$
C_s+\alpha\lambda V_j
\geq\alpha W
\geq\alpha\delta\widehat{n}_{\mathrm{cur},j}
\geq\alpha\delta N_j,
$$
where $N_j$ is the number of active packets at the beginning of the epoch. Any additional successful transmissions can only decrease the number of active packets further.

Consequently, if another estimation epoch follows, its initial number of active packets satisfies:
$$
N_{j+1}\leq(1-\alpha\delta)N_j, 
$$
and thus the number of active packets decreases by a constant factor between successive estimation epochs. Consequently, \mainAlgorithmAcrynom invokes \estimateFunction\ at most $O(\log n)$ times after the initial estimate.

Finally, Lemma~\ref{lem:invariant} guarantees that $\widehat{n}_{\mathrm{cur},j}\leq K N_j$ on event $G$. Since the values $N_j$ decrease geometrically and $N_0\leq n$, we have:
$$
\sum_j\widehat{n}_{\mathrm{cur},j}
\leq K\sum_j N_j
\leq\frac{Kn}{\alpha\delta}
=O(n).
$$
This establishes both claims.
\end{proof}

\begin{lemma}\label{lem:good}
For any fixed $D\geq10$, by choosing sufficiently large constants in the running time of \estimateFunction\ and in the assumption $\rho=\Omega(\log n)$, we obtain:
$$
\Pr(\neg G)=O(n^{-D}).
$$
\end{lemma}
\begin{proof}
We bound the probability that \mainAlgorithmAcrynom obtains an inaccurate population estimate or that a fast round fails the guarantees of Lemma~\ref{lem:exp-number-revised}.

Recall that $G_0$ is the event that the initial call to \estimateFunction\ returns an estimate satisfying $n\leq\widehat{n}\leq32n$ and finishes in $O(\log^2 n)$ slots. On event $G_0$, the initial estimate satisfies $n\leq\widehat{n}\leq32n$. Since \mainAlgorithmAcrynom begins no new fast rounds or calls to \estimateFunction\ after the time cutoff $\tau_{\mathrm{time}}=O(\widehat{n})=O(n)$, at most $O(n)$ fast rounds and subsequent calls to \estimateFunction\ can begin.

Consider any fast round that begins with $r$ active packets satisfying:
$$
\frac{\widehat{n}_{\mathrm{cur}}}{2K}
\leq r\leq\widehat{n}_{\mathrm{cur}}.
$$
By Lemma~\ref{lem:exp-number-revised}, the probability that any subchannel receives too many or too few packets, or that too few packets succeed during an uninterrupted \tstb execution, is at most $O(\lambda e^{-\Omega(x)})
\leq O(\lambda e^{-\Omega(\rho)})$,
since $x\geq\rho$. Because each fast round uses fresh independent random choices, this probability bound holds regardless of the outcomes of preceding rounds.

The analysis of \estimateFunction\ in Appendix~\ref{sec:estimate-n} establishes that each subsequent call returns an accurate estimate with failure probability smaller than any prescribed inverse polynomial in $n$, provided sufficiently many packets remain active. This guarantee holds regardless of the outcomes of earlier calls.

On event $G_0$, \mainAlgorithmAcrynom begins at most $O(n)$ fast rounds and calls to \estimateFunction\ before cleanup. For each fast round satisfying the assumptions of Lemma~\ref{lem:exp-number-revised}, the probability that its guarantees fail is at most $O(\lambda e^{-\Omega(\rho)})$. Similarly, the analysis of \estimateFunction\ in Appendix~\ref{sec:estimate-n} shows that the failure probability of each population estimate can be made smaller than any prescribed inverse polynomial in $n$. Since these guarantees hold regardless of the outcomes of preceding rounds and estimates, we can apply a union bound over all $O(n)$ operations. By choosing the constants in the estimator and in $\rho=\Omega(\log n)$ sufficiently large, the probability that any of these guarantees fails, given $G_0$, is $O(n^{-D})$.

It remains to justify applying Lemma~\ref{lem:exp-number-revised} throughout \mainAlgorithmAcrynom's execution. Lemma~\ref{lem:invariant} establishes that, as long as preceding rounds and population estimates satisfy their guarantees, the number of active packets remains within the range required by Lemma~\ref{lem:exp-number-revised}. Thus, the assumptions of that lemma continue to hold unless one of the failures already accounted for in the union bound occurs.

Finally, the initial population-estimation guarantee established in Appendix~\ref{sec:estimate-n} gives $\Pr(\neg G_0)=O(n^{-D})$, and combining this with the preceding bound yields the claimed upper bound.
\end{proof}

\subsection{Early Cutoff and Fast-Service Accounting}\label{sec:analysis-latency}

For any slow-slot boundary $t$, let $\Service_f([0,t))$ denote the number of fast-channel mini-slots during which the fast channel is available before time $t$. Since \optimalAlgo can transmit at most one packet per fast-channel mini-slot and one packet per slow-channel slot, the total number of packets \optimalAlgo can complete by time $t$ is at most:
$$
\Service([0,t))=t+\Service_f([0,t)).
$$
This bound may include fast-channel mini-slots that \optimalAlgo cannot use because packets can switch channels only at slow-slot boundaries. Nevertheless, since \optimalAlgo completes all $n$ packets by time $T^*=\LatRV(\optimalAlgo,I)$, we must have:
$$
\Service([0,T^*))\geq n.
$$
We next establish an upper bound on $\Service([0,t))$ for times when \mainAlgorithmAcrynom has neither completed all packets nor entered permanent slow-channel cleanup. This bound will allow us to show that \mainAlgorithmAcrynom stops admitting new fast-channel batches by the time \optimalAlgo finishes.

\begin{lemma}
\label{lem:fast-budget}
On event $G$, consider any slow-slot boundary $t<\tau_{\mathrm{time}}$ at which \mainAlgorithmAcrynom has neither completed all packets nor entered permanent slow-channel cleanup. Then:
\begin{equation}\label{eq:service-new}
\Service([0,t))\leq t+O(n/\lambda)
+O\bigl(\rho\log^3(n+2)+F\bigr).
\end{equation}
The constant implicit in $O(n/\lambda)$ is independent of $\lambda$ and the constants used by \estimateFunction.
\end{lemma}
\begin{proof}
Let $A$ be the number of fast rounds that begin by time $t$, and let $h$ be the number of calls to \estimateFunction\ that begin after the initial estimate and by time $t$. Let $M$ be the total number of slots spent executing \estimateFunction\ before time $t$, including the initial call and any call still in progress.

We first bound the number of fast-channel transmission opportunities available to \optimalAlgo while \mainAlgorithmAcrynom executes fast rounds. A round with intended per-subchannel batch size $x$ lasts:
$$
D(x)+2=O(x/\rho)
$$
slow slots, since $x\geq\rho$. During these slots, \optimalAlgo has at most $O(x)$ fast-channel transmission opportunities. Since \mainAlgorithmAcrynom executes only one fast round at a time, summing over all rounds that begin by time $t$ shows that \optimalAlgo has at most $O(\sum x)$ fast-channel transmission opportunities during these rounds. Similarly, the slots spent executing \estimateFunction\ contribute at most $\rho M$ fast-channel transmission opportunities.

We must also account for times when \mainAlgorithmAcrynom is executing \tstb on the slow channel and the fast channel becomes available partway through a slot. Since \mainAlgorithmAcrynom checks fast-channel availability only at slow-slot boundaries, it may miss up to $\rho$ fast-channel mini-slots before responding. The minimum-availability condition ensures that the fast channel remains available at the next boundary, when \mainAlgorithmAcrynom begins a fast round or invokes \estimateFunction.

Each fast round or call to \estimateFunction\ can therefore be preceded by at most one slow-channel slot during which \mainAlgorithmAcrynom misses fast-channel availability. Since $A$ fast rounds and $h$ subsequent calls to \estimateFunction\ have begun by time $t$, these delays account for at most $\rho(A+h)$ additional fast-channel mini-slots. Allowing another $2\rho$ for the beginning and end of the interval being analyzed, we obtain:
$$
\Service_f([0,t))
\leq O\left(\sum x\right)+\rho M+\rho(A+h+2).
$$
Since every fast round has $x\geq\rho$, we have $\rho A\leq\sum x$. Thus, the bound simplifies to:
$$
\Service_f([0,t))
\leq O\left(\sum x\right)+\rho M+O(\rho(h+1)).
$$

It remains to bound the sum of the intended per-subchannel batch sizes. Let $V$ be the sum of $x$ over certified rounds whose reporting slots have finished by time $t$. On event $G$, these rounds complete at least $\alpha\lambda V$ packets. Since there are only $n$ packets, we have:
$$
V\leq\frac{n}{\alpha\lambda}.
$$
Lemma~\ref{lem:charge} bounds the corresponding sum over aborted rounds by $4V+\rho(h+1)$. We also include at most one fast round that has begun but whose outcome has not yet been determined. Its intended per-subchannel batch size is at most $F$. Consequently:
$$
\sum x
\leq5V+\rho(h+1)+F
\leq\frac{5n}{\alpha\lambda}+\rho(h+1)+F.
$$

By Lemma~\ref{lem:epochs}, \mainAlgorithmAcrynom invokes \estimateFunction\ at most $O(\log n)$ times after the initial estimate, so $h=O(\log n)$. Including initialization, the total time spent estimating is $M=O(\log^3(n+2))$ on event $G$. Substituting these bounds into the preceding inequalities gives:
$$
\Service_f([0,t))
\leq O(n/\lambda)+O\bigl(\rho\log^3(n+2)+F\bigr).
$$
Finally, \optimalAlgo can complete at most $t$ packets on the slow channel by time $t$. Therefore:
$$
\Service([0,t))
\leq t+O(n/\lambda)+O\bigl(\rho\log^3(n+2)+F\bigr),
$$
which proves Equation~\ref{eq:service-new}.
\end{proof}

\paragraph{Justifying Our Choice of Constants.}
We now show that the constants required by our analysis can be chosen consistently. First, we choose $\delta>0$ sufficiently small to satisfy the requirements of Lemma~\ref{lem:invariant}. We then choose the fixed fast-channel augmentation factor $\lambda$ sufficiently large so that the $O(n/\lambda)$ term in Equation \ref{eq:service-new} is at most $n/8$.

Next, recall that $F=O(\rho+\sqrt{n/\lambda})$. Since $\rho=O(\sqrt n)$ and $\lambda$ is fixed, we have $\lambda F=O(n/\rho)$. Thus, by choosing $c_{\mathrm{pop}}$ sufficiently large, the population cutoff $\tau_{\mathrm{pop}}=\Theta(n/\rho)$ can satisfy all the requirements in~\eqref{eq:invariant-constants}. We also choose the minimum duration of fast-channel availability sufficiently large for Lemma~\ref{lem:baseline} to apply.

Finally, the constants governing the running time of \estimateFunction\ and the lower bound $\rho=\Omega(\log n)$ are chosen sufficiently large to establish the probability guarantee in Lemma~\ref{lem:good}. Since $\rho=O(\sqrt n)$, the remaining terms in~\eqref{eq:service-new} are at most $n/8$ for sufficiently large $n$. Consequently, under the conditions of Lemma~\ref{lem:fast-budget}, we obtain:
\begin{equation}\label{eq:service-quarter}
\Service([0,t))\leq t+n/4.
\end{equation}

We now show that \mainAlgorithmAcrynom stops admitting new fast-channel batches by the time \optimalAlgo finishes, with high probability. This is important for bounding switching cost, since otherwise \mainAlgorithmAcrynom could continue using the fast channel and incurring switching costs after \optimalAlgo has completed all its packets.

\begin{lemma}
\label{lem:no-post-admission}\label{lem:total-progress-to-opt-time}
Let $T^*=\LatRV(\optimalAlgo,I)$ denote \optimalAlgo's completion time. On event $G$, \mainAlgorithmAcrynom begins no fast round at or after time $T^*$.

If $T^*<\tau_{\mathrm{time}}$, \mainAlgorithmAcrynom has either completed all packets or entered permanent slow-channel cleanup because the population cutoff was reached by time $T^*$. If $T^*\geq\tau_{\mathrm{time}}$, the time cutoff prevents \mainAlgorithmAcrynom from beginning any further fast rounds.
\end{lemma}
\begin{proof}
On $G_0$, for $n\geq2$:
$$ n/128\leq \tau_{\mathrm{time}}=\max\{1,\lfloor \widehat{n}/64\rfloor\}\leq n/2. $$
If $T^*\geq \tau_{\mathrm{time}}$, the protocol's cutoff proves the assertion directly, whether or not a round is currently running.

Suppose $T^*<\tau_{\mathrm{time}}$ and \mainAlgorithmAcrynom has neither completed all packets nor entered permanent slow-channel cleanup by time $T^*$. Initialization takes $O(\log^2 n)=o(n/\rho)$ slots on $G_0$, while $T^*\geq n/(\rho+1)$, so the initial call to \estimateFunction\ has already finished for sufficiently large $n$. Since $T^*<\tau_{\mathrm{time}}\leq n/2$, Lemma~\ref{lem:fast-budget} and Equation~\ref{eq:service-quarter} imply:
$$
n\leq\Service([0,T^*))\leq T^*+n/4<n.
$$
This contradiction shows that \mainAlgorithmAcrynom cannot still be executing before cleanup at time $T^*$. Therefore, \mainAlgorithmAcrynom must have completed all packets or entered permanent slow-channel cleanup by the time \optimalAlgo finishes. Because the time cutoff has not yet been reached, any cleanup must have been triggered by the population cutoff, based on \mainAlgorithmAcrynom's estimate of the remaining active packets. Once cleanup begins, \mainAlgorithmAcrynom never uses the fast channel again.
\end{proof}

\subsection{Cleanup and Worst-Case Bounds}\label{sec:failure}

\begin{lemma}
\label{lem:cleanup}
Suppose \mainAlgorithmAcrynom enters permanent slow-channel cleanup with $r$ active packets and uses an estimate $v$ to determine the initial \tstb size parameter. Then, regardless of the accuracy of $v$, the expected time until all remaining packets succeed is $O(r+v)$. Cleanup incurs no switching cost.
\end{lemma}
\begin{proof}
\mainAlgorithmAcrynom begins cleanup with a \tstb execution using size parameter $s_0=\max\{v,s_{\min}\}$, where $s_{\min}$ is a sufficiently large fixed constant. If packets remain unsuccessful after this execution, \mainAlgorithmAcrynom begins another with twice the size parameter. This continues until all packets succeed.

First, consider the executions whose size parameters are smaller than $r$, the number of active packets when cleanup begins. Since the parameters double after each execution and \tstb with parameter $s$ takes $O(s)$ slow-channel slots, the total time spent on these executions is $O(r)$.

Once the size parameter $s$ reaches or exceeds $r$, there are at most $s$ active packets. The \tstb completion guarantee implies that all remaining packets succeed with probability at least $1-O(s^{-d})$. This guarantee applies even when fewer than $s$ packets participate, because removing competing packets cannot prevent a successful transmission. By choosing $s_{\min}$ sufficiently large, the probability that any packets remain unsuccessful after such an execution is at most $1/8$.

Let $s$ now denote the size parameter of the first execution for which $s\geq r$. Since the parameters double, we have $s=O(r+v)$. Each subsequent execution also doubles the parameter, but occurs only if the preceding execution fails to finish all remaining packets. Thus, the expected time spent on these executions is at most:
$$
O(s)\sum_{j\geq0}2^j(1/8)^j
=O(s)
=O(r+v).
$$
Combining this with the time spent on earlier executions gives expected cleanup time $O(r+v)$.

Finally, all transmissions during cleanup occur on the slow channel, so no switching cost is incurred.
\end{proof}

The preceding analysis relies on high-probability guarantees for population estimation and fast-channel progress. To establish our expected performance bounds, we must also account for executions in which these guarantees fail. The following lemma provides upper bounds that hold regardless of whether those guarantees are satisfied.

\begin{lemma}
\label{lem:safe}
Let $\tau_0$ be the time spent obtaining the initial population estimate. On every execution, \mainAlgorithmAcrynom either finishes or enters permanent slow-channel cleanup within $\tau_0+O(\widehat{n})$ slots, and its switching cost is $O(n\widehat{n})$.

Once cleanup begins, the expected time needed to finish the remaining packets is $O(n+\widehat{n})$, regardless of the accuracy of the population estimates.
\end{lemma}
\begin{proof}
\mainAlgorithmAcrynom does not begin any new fast rounds at or after the time cutoff $\tau_{\mathrm{time}}=O(\widehat{n})$. Since at most one fast round can begin in any slow-channel slot, \mainAlgorithmAcrynom begins at most $O(\widehat{n})$ fast rounds before this cutoff.

Each fast round has intended per-subchannel batch size $x\leq\widehat{n}/\lambda$ and lasts:
$$
D(x)+2=O(\widehat{n}/\rho+1)=O(\widehat{n})
$$
slow slots, including the two slots used to determine whether an eviction occurred. If a round is still underway when the time cutoff is reached, \mainAlgorithmAcrynom finishes that round before entering permanent slow-channel cleanup. When \mainAlgorithmAcrynom is executing \tstb on the slow channel or running \estimateFunction, it checks the time cutoff at every slow-slot boundary. If the initial call to \estimateFunction\ finishes after the cutoff, \mainAlgorithmAcrynom enters cleanup immediately.

Consequently, regardless of the accuracy of its population estimates, \mainAlgorithmAcrynom either completes all packets or enters cleanup within $\tau_0+O(\widehat{n})$ slots.

We next bound switching cost. Each fast round moves at most one batch onto the fast channel, incurring one unit of cost. At most $n$ packets can return to the slow channel during that round, and each packet returns at most once, whether because of an eviction or because it has not succeeded by the scheduled end of the round. Since \mainAlgorithmAcrynom begins at most $O(\widehat{n})$ fast rounds, its total switching cost is:
$$
O(n\widehat{n}).
$$

Finally, when \mainAlgorithmAcrynom enters cleanup, the initial \tstb size parameter is determined by either $\widehat{n}_{\mathrm{cur}}$, if the population cutoff is reached, or $\widehat{n}$, if the time cutoff is reached. Both are at most $\widehat{n}$. Since at most $n$ packets remain active, Lemma~\ref{lem:cleanup} guarantees that their expected completion time is:
$$
O(n+\widehat{n}).
$$
This establishes all three bounds.
\end{proof}

\begin{proposition}
\label{lem:latency-constant-factor}
For every fixed instance $I$, \mainAlgorithmAcrynom achieves expected latency:
$$
E[\LatRV(\mainAlgorithmAcrynom,I)]=O(\LatRV(\optimalAlgo,I)).
$$
\end{proposition}

\begin{proof}
Let $T^*=\LatRV(\optimalAlgo,I)$ denote \optimalAlgo's completion time. We first establish the bound on event $G$, when the population estimates and fast-channel executions satisfy their high-probability guarantees. We consider two cases, depending on when \optimalAlgo finishes.

\medskip
\noindent{\it Case 1: \optimalAlgo finishes before the time cutoff.}
Suppose $T^*<\tau_{\mathrm{time}}$. By Lemma~\ref{lem:no-post-admission}, \mainAlgorithmAcrynom has either completed all packets or entered permanent slow-channel cleanup by time $T^*$. Since the time cutoff has not been reached, cleanup must have been triggered by the population cutoff.

On event $G$, the number $r$ of remaining active packets satisfies:
$$
r\leq\widehat{n}_{\mathrm{cur}}
\leq\tau_{\mathrm{pop}}
=O(n/\rho).
$$
By Lemma~\ref{lem:cleanup}, the expected time needed to finish these packets on the slow channel is $O(n/\rho)$. Since \optimalAlgo requires at least $\Omega(n/\rho)$ slots to transmit all $n$ packets, \mainAlgorithmAcrynom's expected latency in this case is $O(T^*)$.

\medskip
\noindent{\it Case 2: \optimalAlgo finishes at or after the time cutoff.}
Suppose $T^*\geq\tau_{\mathrm{time}}$. On event $G$, the initial estimate satisfies $n\leq\widehat{n}\leq32n$, so:
$$
T^*\geq\tau_{\mathrm{time}}\geq n/128.
$$
Thus $T^*=\Omega(n)$. Lemma~\ref{lem:safe} guarantees that \mainAlgorithmAcrynom completes all packets or enters permanent slow-channel cleanup within $O(n)$ slots. By Lemma~\ref{lem:cleanup}, finishing any remaining packets takes another $O(n)$ expected slots. Therefore, \mainAlgorithmAcrynom's expected latency is again $O(T^*)$.

\medskip
\noindent{\it Executions in which the guarantees fail.}
It remains to account for executions outside event $G$. Recall that $G_0$ concerns the accuracy and running time of the initial population estimate, while $G$ also includes the guarantees for subsequent estimates and fast rounds.

If $G_0$ holds but $G$ fails, Lemmas~\ref{lem:good} and~\ref{lem:safe} show that this event has probability $O(n^{-D})$ and that \mainAlgorithmAcrynom's expected latency is bounded by $O(n)$, even when these guarantees fail. Its contribution to the overall expectation is therefore $O(n^{1-D})$.

If $G_0$ fails, Lemma~\ref{lem:safe} bounds the expected latency by $\tau_0+O(n+\widehat n)$. By Lemma~\ref{lem:init}, the contributions of $\tau_0$ and $\widehat n$ from these executions are each $O(n^{-D})$, while the fixed $n$ term contributes $O(n\Pr(\neg G_0))=O(n^{1-D})$. Thus, these executions also contribute at most $O(n^{1-D})$ to expected latency.

These contributions are negligible compared with $T^*=\Omega(n/\rho)$. Combining them with the bounds established on $G$ gives $E[\LatRV(\mainAlgorithmAcrynom,I)]=O(T^*)$, 
as claimed.
\end{proof}

\subsection{Switching Cost}\label{sec:analysis-switching-cost}

On event $G$, \mainAlgorithmAcrynom begins no new fast rounds after \optimalAlgo finishes at time $T^*$. However, a round that began before $T^*$ may finish afterward, potentially incurring additional costs when unsuccessful packets return to the slow channel. Our analysis needs to include these costs.

The following lemma establishes that \optimalAlgo cannot ignore more than two iterations in which it could have used the fast channel before finishing. This allows us to bound the number of iterations in which \mainAlgorithmAcrynom may incur eviction costs in terms of the number of iterations actually used by \optimalAlgo.

\begin{lemma}\label{lem:offline}
Let $M$ be the number of iterations containing a slow-slot boundary before $T^*$ at which packets could enter the fast channel. If \optimalAlgo successfully transmits packets on the fast channel in $\TNI$ iterations, then $M\leq\TNI+2$.
\end{lemma}
\begin{proof}
For each integer $t$, let $C(t)$ denote the total number of transmission opportunities available to \optimalAlgo before time $t$. This includes one opportunity per slow-channel slot, even during eviction periods, and one per fast mini-slot beginning at the first slow-slot boundary in each iteration when packets can enter the fast channel. Since all packets are present initially, \optimalAlgo can use these opportunities to transmit distinct packets without collisions, assigning packets to the fast channel at the first permitted boundary of each iteration. Consequently, $T^*$ is the smallest integer $t$ for which $C(t)\geq n$.

Since $C(T^*-1)\leq n-1$ and one slow slot adds at most $\rho+1$ opportunities,
$$ 0\leq C(T^*)-n\leq\rho. $$
Every iteration that ends before $T^*$ provides at least $\rho$ usable fast-channel transmission opportunities. Since \optimalAlgo can leave at most $\rho$ opportunities unused by time $T^*$, it cannot ignore two such iterations without contradicting the preceding bound. At most one additional iteration can still be underway at $T^*$. Consequently, $M\leq\TNI+2$ for any minimum-latency schedule.
\end{proof}

\begin{lemma}\label{lem:pathcost}\label{lem:large-iterations}
On event $G$, for $\TNI\geq1$, \mainAlgorithmAcrynom executes at most $O(\TNI\rho+\sqrt{\TNI n})$ fast rounds. The total switching cost from admitting these batches and returning packets evicted before their scheduled deadlines is also $O(\TNI\rho+\sqrt{\TNI n})$.
\end{lemma}
\begin{proof}
Recall that $M$ counts the iterations in which packets could enter the fast channel before $T^*$. By Lemma~\ref{lem:no-post-admission}, every fast round begins before $T^*$. Each aborted round begins during an iteration that ends with the eviction responsible for its abort. No two aborted rounds can begin in the same iteration, since the eviction interrupting the first round ends that iteration before another round can begin. This remains true even if the eviction or its report occurs after $T^*$. Thus, the number $J$ of aborted rounds satisfies $J\leq M\leq\TNI+2$.

Each reported abort resets $b$ to zero, whereas a new population estimate leaves $b$ unchanged. Let $t_1,\ldots,t_{J+1}$ denote the numbers of certified rounds before the first reported abort, between successive reported aborts, and after the last reported abort, respectively. Since the population cap is inactive on $G$, the intended number of packets per subchannel in each such sequence increases through $\rho,\rho+1,\ldots,\rho+t_i-1$.

On $G$, these certified rounds complete at least $\alpha\lambda x$ packets each, where $x$ is the intended number of packets per subchannel. Since $J$ aborted rounds divide the execution into at most $J+1$ sequences of certified rounds, Cauchy--Schwarz gives the following bound on their total number:
$$
\frac{\alpha\lambda}{2}\sum_i t_i^2\leq n,
\qquad
\sum_i t_i\leq\sqrt{\frac{2(J+1)n}{\alpha\lambda}}.
$$

There are at most $\sum_i t_i+J+1$ fast rounds in total, including a possible final round in which all remaining packets succeed before its outcome is recorded. Each round incurs one batch-admission cost.

An aborted round ending the $i$-th sequence has intended per-subchannel batch size $\rho+t_i$. On $G$, at most $2\lambda(\rho+t_i)$ packets participate in that round and can therefore be returned by its eviction. A round in which the last packet succeeds during reporting is included among the aborted rounds.

The total cost of moving batches onto the fast channel is at most $\sum_i t_i+J+1$, since each round incurs one unit of cost. The cost of returning packets evicted before their scheduled deadlines is at most $2\lambda J\rho+2\lambda\sum_i t_i$. Combining these bounds gives:
$$
O\left(\lambda(J+1)\rho+\lambda\sum_i t_i\right)
=O\left(\lambda(J+1)\rho+
\sqrt{\lambda(J+1)n}\right).
$$
The equality follows from the preceding bound on $\sum_i t_i$. Since $J\leq\TNI+2$, $\lambda$ is constant, and $\TNI\geq1$, the total is $O(\TNI\rho+\sqrt{\TNI n})$.
\end{proof}
\begin{lemma}\label{lem:expectedcost}
For a fixed instance with $\TNI\geq1$, let $C_{\mathrm{ret}}$ denote the total cost of returning packets that remain on the fast channel at the scheduled end of a round's \tstb execution. Then:
$$
E[C_{\mathrm{ret}}\mid G]\Pr(G)
=O(\TNI\rho+\sqrt{\TNI n}).
$$
\end{lemma}
\begin{proof}
By Lemma~\ref{lem:pathcost}, on event $G$, the number of fast rounds is at most $A_{\max}=O(\TNI\rho+\sqrt{\TNI n})$, where $A_{\max}$ is a fixed upper bound.

We consider the fast rounds in the order they begin. For each $j\leq A_{\max}$, let $R_j$ denote the number of packets that return to the slow channel at the scheduled end of the $j$-th round, provided that round begins with at most $\widehat n_{\mathrm{cur}}$ active packets. Otherwise, including when the $j$-th round does not occur, we set $R_j=0$.

Consider any round that begins with at most $\widehat n_{\mathrm{cur}}$ active packets. By Lemma~\ref{lem:return}, the expected number of packets remaining unsuccessful after its full scheduled \tstb execution, assuming no eviction, is $O(1)$ per logical subchannel, or $O(\lambda)$ across all subchannels.

If an eviction occurs before the scheduled end of the round, unfinished packets return at the time of eviction rather than at the scheduled end. Their switching costs are accounted for in Lemma~\ref{lem:pathcost}. Otherwise, the number returning at the scheduled end is bounded by the number that would remain unsuccessful after the full \tstb execution.

Consequently, the expected number of packets returning at the scheduled end of any such round is $O(\lambda)$. This bound holds regardless of the outcomes of earlier rounds, since each round uses fresh independent random choices. In particular, $E[R_j]=O(\lambda)$ for every $j\leq A_{\max}$.

On event $G$, every fast round begins with at most $\widehat n_{\mathrm{cur}}$ active packets, and there are at most $A_{\max}$ rounds. Therefore:
$$
\begin{aligned}
E[C_{\mathrm{ret}}\mid G]\Pr(G)
&\leq\sum_{j=1}^{A_{\max}}E[R_j]\\
&\leq O(\lambda A_{\max})\\
&=O(\TNI\rho+\sqrt{\TNI n}),
\end{aligned}
$$
where the last equality uses the fact that $\lambda$ is constant. 
\end{proof}

\begin{proposition}\label{lem:revised-switching-cost}\label{lem:switching-cost-total}
For $\TNI\geq1$, the expected switching cost of \mainAlgorithmAcrynom satisfies:
$$
E[\Cost(\mainAlgorithmAcrynom,I)]
=O(\TNI\rho+\sqrt{\TNI n}),
$$
and its switching-cost competitive ratio is:
$$
\frac{E[\Cost(\mainAlgorithmAcrynom,I)]}{\Cost(\optimalAlgo,I)}
=O\left(\rho+\sqrt{n/\TNI}\right).
$$
For $\TNI=0$, \mainAlgorithmAcrynom incurs zero switching cost on $G_0$. Moreover, by choosing the estimator's confidence constant sufficiently large, its expected switching cost can be made polynomially small in $n$.
\end{proposition}
\begin{proof}
First suppose $\TNI\geq1$. On event $G$, Lemma~\ref{lem:pathcost} bounds the cost of moving batches onto the fast channel and returning packets evicted before their scheduled fast-channel executions finish. Lemma~\ref{lem:expectedcost} bounds the expected cost of returning packets that remain unsuccessful at the scheduled end of these executions. Together, these lemmas account for all switching costs, including those incurred by rounds that begin before $T^*$ but finish afterward. Consequently, on $G$, the expected switching cost is $O(\TNI\rho+\sqrt{\TNI n})$.

We must also account for executions in which $G$ fails. When $G_0$ holds but $G$ does not, Lemma~\ref{lem:safe} bounds the switching cost by $O(n\widehat n)=O(n^2)$. Lemma~\ref{lem:good} guarantees that these executions occur with sufficiently small probability that their contribution to the expected switching cost is negligible. When $G_0$ fails, Lemma~\ref{lem:safe} again bounds the switching cost by $O(n\widehat n)$, while Lemma~\ref{lem:init} guarantees that the expected contribution of these executions is also negligible. Since the constants in the estimator can be chosen to make these contributions polynomially small in $n$, the overall expected switching cost remains $O(\TNI\rho+\sqrt{\TNI n})$.

To obtain the competitive ratio, recall that all packets are initially on the slow channel. In each of the $\TNI$ iterations in which \optimalAlgo successfully transmits packets on the fast channel, it must move at least one batch onto that channel, incurring at least one unit of switching cost. Therefore, $\Cost(\optimalAlgo,I)\geq\TNI$, and the claimed competitive ratio follows.

Now suppose $\TNI=0$. Since \optimalAlgo never successfully transmits on the fast channel, its minimum completion time is $T^*=n$, achieved by transmitting all packets on the slow channel. Moreover, the fast channel cannot be available at any slow-slot boundary $t\leq n-2$. Otherwise, \optimalAlgo could complete one packet on the fast channel and the remaining $n-1$ packets on the slow channel, finishing by time $n-1$. This would contradict the minimum completion time $T^*=n$.

On $G_0$, the initial estimate satisfies $\widehat n\leq32n$, so the time cutoff satisfies $\tau_{\mathrm{time}}\leq n/2\leq n-1$. Since \mainAlgorithmAcrynom can begin a fast round only at a slow-slot boundary before this cutoff, and the fast channel is unavailable at every such boundary, no fast round can begin. Thus, \mainAlgorithmAcrynom remains on the slow channel and incurs zero switching cost on $G_0$.

Finally, Lemma~\ref{lem:safe} bounds the switching cost on every execution by $O(n\widehat n)$, while Lemma~\ref{lem:init} guarantees that executions outside $G_0$ contribute a polynomially small amount to its expected value. This establishes the claimed expected switching cost when $\TNI=0$. No switching-cost competitive ratio is asserted in this case, since \optimalAlgo incurs zero switching cost. If the fast channel is never available, \mainAlgorithmAcrynom also incurs zero switching cost on every execution.
\end{proof}

\begin{proof}[Proof of Theorem~\ref{thm:main-theorem}]
As described in Section~\ref{sec:protocol}, \mainAlgorithmAcrynom is decentralized: packets use common channel feedback and synchronized slot boundaries to coordinate their actions without centralized control. Proposition~\ref{lem:latency-constant-factor} establishes that \mainAlgorithmAcrynom achieves expected latency $O(T^*)$, while Proposition~\ref{lem:switching-cost-total} establishes the claimed switching-cost bound. Finally, population estimation, slow-channel contention resolution, eviction reporting, and cleanup all use at most one slow-channel transmission opportunity per slot. Thus, resource augmentation is required only on the fast channel.
\end{proof}

\section{Lower Bounds}\label{sec:analysis-lower-bounds}

Our lower bounds hold even when the online algorithm is given constant-factor resource augmentation on both the slow and fast channels (in contrast, our upper bound assumes augmentation only on the fast channel). We establish each bound by constructing an eviction schedule that forces substantial switching cost on any randomized online algorithm achieving constant-competitive expected latency.

\subsection{Switching-Cost Ratio}

To prove Theorem~\ref{thm:lower-bounds}(A), let $c\geq1$ be the constant in the assumed expected-latency guarantee, and let $\lambda_s$ and $\lambda_f$ denote the fixed augmentation factors on the slow and fast channels. We construct a distribution over eviction schedules and first establish a switching-cost lower bound for deterministic algorithms that finish the no-eviction instance quickly. We then extend the result to randomized algorithms.

Let:
$$
L=\lceil n/\rho\rceil+1,\qquad K_0=\lceil4cL\rceil.
$$
Thus, $K_0=\Theta(n/\rho)$. Choose a fixed integer $h_0\geq C_{\mathrm{avail}}$, and consider the set of times:
$$
\mathcal T=\left\{u+j/\rho:
 h_0\leq u<K_0,\ 1\leq j<\rho\right\}.
$$
Each time in $\mathcal T$ is a fast-channel mini-slot boundary strictly inside a slow-channel slot. An eviction beginning at any such time is permitted, because the fast channel has already been available for at least $C_{\mathrm{avail}}$ slots. There are $K_0-h_0$ slow-channel slots between $h_0$ and $K_0$, each containing $\rho-1$ such boundaries. Consequently:
$$
|\mathcal T|=(K_0-h_0)(\rho-1)
=\Theta(n/\rho)\cdot\Theta(\rho)=\Theta(n).
$$

We define a probability distribution $\mu$ over eviction schedules. With probability $1/2$, the fast channel remains available throughout the execution; call this instance $I_0$. Otherwise, a time $t$ is chosen uniformly from $\mathcal T$, and the only eviction occupies the single fast mini-slot $[t,t+1/\rho)$; call this instance $I_t$. We write $E_{I\sim\mu}$ for expectation over the choice of an instance from this distribution.

\optimalAlgo knows the eviction schedule and can ensure that all packets assigned to the fast channel succeed before an eviction. After the one-mini-slot eviction, \optimalAlgo may have to wait until the next slow-slot boundary before assigning another batch to the fast channel, losing at most one slow slot of fast-channel capacity. \optimalAlgo can therefore finish within $L$ slots, using at most two batches on the fast channel and incurring switching cost $O(1)$. Moreover, a schedule achieving minimum latency can be chosen with this switching cost: packets are assigned to the earliest usable slow- and fast-channel transmission slots, and each fast-channel batch enters at the first permitted slow-slot boundary of its availability interval.

\begin{lemma}
\label{lem:lb-deterministic}
For the distribution $\mu$ defined above, every deterministic online algorithm $\mathcal A$ that completes all packets on $I_0$ within $K_0$ slow-channel slots satisfies:
$$
E_{I\sim\mu}[\Cost(\mathcal A,I)]=\Omega(\sqrt n).
$$
\end{lemma}

\begin{proof}
Fix such a deterministic algorithm $\mathcal A$. By time $K_0$, it can complete at most $\lambda_sK_0=o(n)$ packets on the slow channel, and at most $\lambda_f\rho h_0=o(n)$ packets on the fast channel before time $h_0$. Since it finishes all $n$ packets on $I_0$ by $K_0$, at least $\Omega(n)$ packets succeed on the fast channel between times $h_0$ and $K_0$.

We group these packets according to their last move onto the fast channel before succeeding. Packets whose last move occurred at or before $h_0$ form one group, which begins at time $h_0$. For each batch moved onto the fast channel after $h_0$, the packets for which that batch was their last move form another group, beginning when the batch enters the fast channel. Only packets that succeed on the fast channel during $[h_0,K_0)$ are included. Let $s_1,\ldots,s_g$ be the sizes of the nonempty groups. Then:
$$
S=\sum_{i=1}^g s_i=\Omega(n).
$$
Each group other than the possible initial group corresponds to a distinct batch moved onto the fast channel, and a nonempty initial group requires at least one earlier batch. Thus, the switching cost on $I_0$ is at least $g$.

If $g\geq S/(8\lambda_f)$, the cost of moving these batches onto the fast channel is already $\Omega(n)$, which proves a stronger bound. Otherwise $g<{S}/{8\lambda_f}$. The groups with fewer than $4\lambda_f$ packets account for fewer than $4\lambda_f g<S/2$ packets. Therefore, ``larger'' groups with $s_i\geq4\lambda_f$ together contain at least $S/2=\Omega(n)$ packets.

Consider one of these larger groups. By the definition of the groups, its $s_i$ packets remain on the fast channel from the group's starting time until they succeed. The augmented fast channel can complete at most $\lambda_f$ packets per mini-slot. Hence, during the first $\lfloor s_i/(2\lambda_f)\rfloor$ mini-slot boundaries after the group begins, at least $s_i/2$ packets from the group remain on the fast channel.

Recall that each group contains only packets that successfully transmit on the fast channel before time $K_0$. Consequently, the mini-slot boundaries under consideration must occur before $K_0$, since transmitting all $s_i$ packets requires at least $s_i/\lambda_f$ mini-slots. At least half of these boundaries lie strictly inside slow-channel slots and therefore belong to $\mathcal T$. Since at least $s_i/2$ packets from the group remain on the fast channel at each of these times, the sum of the number of packets exposed to eviction over these times is at least:
$$
\Omega\left(\frac{s_i^2}{\lambda_f}\right)=\Omega(s_i^2).
$$

For each $t\in\mathcal T$, let $x_t$ be the number of packets on the fast channel at time $t$ in the execution of $\mathcal A$ on $I_0$, after any transmissions ending at $t$ have completed. Summing over the larger groups gives:
$$
\sum_{t\in\mathcal T}x_t
\geq\Omega\left(\sum_{i:s_i\geq4\lambda_f}s_i^2\right).
$$
Recall that the groups with $s_i\geq4\lambda_f$ together contain at least $S/2=\Omega(n)$ packets. By Cauchy--Schwarz, the square of the sum of their sizes is at most the number of these groups multiplied by the sum of their squared sizes. Since there are at most $g$ such groups, we have:
$$
\left(\sum_{i:s_i\geq4\lambda_f}s_i\right)^2
\leq g\sum_{i:s_i\geq4\lambda_f}s_i^2.
$$
Rearranging, and using the fact that the sum on the left is $\Omega(n)$, gives:
$$
\sum_{i:s_i\geq4\lambda_f}s_i^2
\geq\frac{\left(\sum_{i:s_i\geq4\lambda_f}s_i\right)^2}{g}
=\Omega(n^2/g).
$$
Therefore:
$$
\sum_{t\in\mathcal T}x_t=\Omega(n^2/g).
$$

On instance $I_t$, the execution agrees with the execution on $I_0$ until time $t$. Because $t$ lies strictly inside a slow-channel slot, no new packet can move onto the fast channel at that time. The eviction therefore forces all $x_t$ packets then on the fast channel to return to the slow channel.

The distribution $\mu$ assigns probability $1/2$ to $I_0$ and probability $1/(2|\mathcal T|)$ to each $I_t$. Combining the cost of moving batches onto the fast channel on $I_0$ with the expected eviction cost gives:
$$
\begin{aligned}
E_{I\sim\mu}[\Cost(\mathcal A,I)]
&\geq \frac g2+\frac{1}{2|\mathcal T|}\sum_{t\in\mathcal T}x_t\\
&\geq\Omega(g)+\Omega(n/g)\\
&=\Omega(\sqrt n),
\end{aligned}
$$
where $|\mathcal T|=\Theta(n)$. If $g\geq S/(8\lambda_f)$, the same conclusion follows directly from the switching cost on $I_0$. Thus the lemma holds in both cases.
\end{proof}

The preceding argument resembles Yao's minimax principle, but requires an additional step because we consider randomized algorithms that are constant-competitive in expected latency. A randomized algorithm can be viewed as a distribution over deterministic algorithms, some of which may not be latency-competitive even when the randomized algorithm is competitive in expected latency. Consequently, a switching-cost lower bound proved only for latency-competitive deterministic algorithms would not necessarily extend to randomized algorithms whose distributions include deterministic algorithms that are not latency-competitive.

To avoid this difficulty, our deterministic lower bound requires only that the algorithm finish the no-eviction instance $I_0$ within $K_0=O(n/\rho)$ slots. As we show next, Markov's inequality guarantees that, for any randomized algorithm satisfying our expected-latency requirement, a constant fraction of its probability distribution is over deterministic algorithms that finish $I_0$ within $K_0$ slots. We can then apply the deterministic lower bound to those algorithms and average to obtain the desired result.

\begin{lemma}
\label{lem:lb-randomized}
For the distribution $\mu$, every randomized online algorithm $\mathcal R$ that is $c$-competitive in expected latency incurs expected switching cost $\Omega(\sqrt n)$, where the expectation is over both the algorithm's random choices and the eviction schedule drawn from $\mu$.
\end{lemma}
\begin{proof}
Since \optimalAlgo finishes the no-eviction instance $I_0$ within $L$ slow-channel slots and $\mathcal R$ is $c$-competitive in expected latency, we have:
$$
E[\LatRV(\mathcal R,I_0)]
\leq c\LatRV(\optimalAlgo,I_0)\leq cL.
$$
Recall that $K_0=\lceil4cL\rceil$. By Markov's inequality, the probability that $\mathcal R$ takes more than $K_0$ slots to finish $I_0$ is at most:
$$
\Pr\bigl(\LatRV(\mathcal R,I_0)>K_0\bigr)
\leq\frac{E[\LatRV(\mathcal R,I_0)]}{K_0}
\leq\frac{cL}{4cL}
=\frac14.
$$
Therefore, with probability at least $3/4$ over the random choices of $\mathcal R$, the algorithm finishes $I_0$ within $K_0$ slots.

Recall that a randomized algorithm can be viewed as a distribution over deterministic algorithms. For each choice of randomness that results in completion within $K_0$ slots, Lemma~\ref{lem:lb-deterministic} guarantees expected switching cost $\Omega(\sqrt n)$ when the eviction schedule is drawn from $\mu$. 

These choices account for at least $3/4$ of the probability distribution over deterministic algorithms. Since switching costs are nonnegative for all other choices, averaging over the random choices of $\mathcal R$ and the eviction schedules drawn from $\mu$ gives expected switching cost $\Omega(\sqrt n)$, as claimed.
\end{proof}

\begin{proof}[Proof of Theorem~\ref{thm:lower-bounds}(A)]
By Lemma~\ref{lem:lb-randomized}, the average expected switching cost of $\mathcal R$ over the finitely many instances in $\mu$ is $\Omega(\sqrt n)$. Thus, some fixed instance in this distribution gives expected switching cost $\Omega(\sqrt n)$. As established above, \optimalAlgo achieves minimum latency with switching cost $O(1)$ on every instance in the distribution. Hence this fixed instance gives an expected switching-cost competitive ratio $\Omega(\sqrt n)$, as claimed.
\end{proof}

\subsection{A Rate-Dependent Switching-Cost Lower Bound}

The key idea behind the proof below is that an algorithm with constant-competitive expected latency must finish in $O(n/\rho)$ slots with constant probability when the fast channel is continuously available. Since the slow channel can complete only $o(n)$ packets in that time, most packets must finish on the fast channel. Because packets can enter the fast channel only at slow-slot boundaries, this requires $\Omega(\rho)$ packets to remain on the fast channel after the first mini-slot in a constant fraction of slow slots. An eviction at a randomly chosen such time therefore incurs expected cost $\Omega(\rho)$. Since \optimalAlgo can achieve minimum latency with unit switching cost on every instance in the construction, an averaging argument yields a fixed eviction schedule with expected switching-cost competitive ratio $\Omega(\rho)$.

\begin{proof}[Proof of Theorem~\ref{thm:lower-bounds}(B)]
Fix a randomized online algorithm $\mathcal R$ that is $c$-competitive in expected latency. Let $\lambda_s$ and $\lambda_f$ be its fixed augmentation factors on the slow and fast channels, respectively. Choose a fixed integer $h_0\geq C_{\mathrm{avail}}$, and let:
$$
L=\lceil n/(\rho+1)\rceil,\qquad K_0=\lceil4cL\rceil.
$$
On the instance $I_0$, where the fast channel is always available, \optimalAlgo finishes within $L$ slots. As in the preceding lower bound, Markov's inequality implies that $\mathcal R$ finishes $I_0$ within $K_0$ slots with probability at least $3/4$. We fix any choice of the algorithm's random decisions for which this occurs and consider the resulting deterministic execution.

For each integer $u\in\{h_0,\ldots,K_0-1\}$, consider an eviction beginning at time $u+1/\rho$, just after the first mini-slot of slow-channel slot $u$ ends. The fast channel remains unavailable thereafter. Let $I_u$ denote the resulting instance. Each such eviction is permitted because the fast channel has been available for at least $C_{\mathrm{avail}}$ slots.

On every instance $I_u$, \optimalAlgo can achieve minimum latency with switching cost exactly one. Indeed, any minimum-latency schedule can be modified so that all packets that succeed on the fast channel enter it together at time zero, while all other packets remain on the slow channel. Keeping only the original successful transmissions preserves their feasibility and completion times: all packets are present initially, and the fast channel is continuously available until the sole eviction. All fast-channel packets succeed before the eviction takes effect, including any transmission ending at the eviction boundary, so no packet returns to the slow channel. Furthermore, a schedule using only the slow channel requires $n$ slots, whereas sending one packet in the first fast mini-slot and the other $n-1$ packets on the slow channel finishes by time $n-1$. Thus, every minimum-latency schedule uses the fast channel. Choosing a minimum-switching schedule among the minimum-latency schedules therefore gives $\Cost(\optimalAlgo,I_u)=1$.

Let $q_u$ be the number of packets remaining on the fast channel immediately after that first mini-slot in the execution on $I_0$. We will show that $q_u=\Omega(\rho)$ for a constant fraction of the slow-channel slots under consideration.

Let $\mathcal H$ denote the set of slow-channel slots in which at least $a\rho$ packets remain on the fast channel immediately after the first mini-slot, where $a>0$ is a sufficiently small fixed constant. That is:
$$
\mathcal H=\{u\in\{h_0,\ldots,K_0-1\}:q_u\geq a\rho\}.
$$

Consider a slow-channel slot $u\notin\mathcal H$ with $u\geq h_0$. At most $\lambda_f$ packets can succeed during its first fast mini-slot. Since packets can move onto the fast channel only at slow-slot boundaries, no additional packets can enter the fast channel during the remainder of that slot. Since $q_u<a\rho$, at most another $a\rho$ packets can succeed on the fast channel during this slot. Thus, such a slot contributes at most $\lambda_f+a\rho$ fast-channel successes. For slots in $\mathcal H$, we use the upper bound of $\lambda_f\rho$ successful fast-channel transmissions.

We now count the successful transmissions by time $K_0$. The slow channel completes at most $\lambda_sK_0$ packets. On the fast channel, the first $h_0$ slots contribute at most $h_0\lambda_f\rho$ successes, the $|\mathcal H|$ slots in $\mathcal H$ contribute at most $|\mathcal H|\lambda_f\rho$, and the remaining $K_0-h_0-|\mathcal H|$ slots contribute at most $(K_0-h_0-|\mathcal H|)(\lambda_f+a\rho)$. Since all $n$ packets succeed by $K_0$, we obtain:
$$
n\leq \lambda_sK_0+h_0\lambda_f\rho
+|\mathcal H|\lambda_f\rho
+(K_0-h_0-|\mathcal H|)(\lambda_f+a\rho).
$$
In particular:
$$
n\leq \lambda_sK_0+h_0\lambda_f\rho
+|\mathcal H|\lambda_f\rho+K_0(\lambda_f+a\rho).
$$

Recall that $K_0=\Theta(n/\rho)$. Since $\rho=\Omega(\log n)$ and $\rho=O(\sqrt n)$, and since $h_0$, $\lambda_s$, and $\lambda_f$ are constants, the terms $\lambda_sK_0$, $h_0\lambda_f\rho$, and $\lambda_fK_0$ are all $o(n)$. We choose $a$ sufficiently small that $K_0a\rho\leq n/4$ for sufficiently large $n$. Therefore, $|\mathcal H|\lambda_f\rho=\Omega(n)$, which implies:
$$
|\mathcal H|=\Omega(n/\rho)=\Omega(K_0).
$$
Thus, during a constant fraction of the slow-channel slots between $h_0$ and $K_0$, at least $a\rho$ packets remain on the fast channel immediately after the first mini-slot.

Now consider the execution on $I_u$ for some $u\in\mathcal H$. Since the eviction occurs immediately after the first fast mini-slot, the execution on $I_u$ agrees with that on $I_0$ up to that time. In particular, the same packets are on the fast channel when the eviction occurs. Since $q_u\geq a\rho$, the eviction forces at least $a\rho$ packets to return to the slow channel.

Suppose we choose $u$ uniformly at random from $\{h_0,\ldots,K_0-1\}$. Because $\mathcal H$ contains a constant fraction of these choices, the expected eviction cost is $\Omega(\rho)$ for every fixed choice of the algorithm's random decisions that finishes $I_0$ within $K_0$ slots.

As established above, such choices of the algorithm's randomness occur with probability at least $3/4$. Averaging over all random choices therefore gives expected eviction cost $\Omega(\rho)$, where the expectation is over both the algorithm's randomness and the uniformly chosen eviction time.

Since the average expected eviction cost over the candidate schedules is $\Omega(\rho)$, at least one fixed schedule $I_u$ must incur expected eviction cost $\Omega(\rho)$, and hence expected switching cost $\Omega(\rho)$, where the expectation is only over the random choices of $\mathcal R$. Since $\Cost(\optimalAlgo,I_u)=1$, this fixed instance satisfies:
$$
\frac{E[\Cost(\mathcal R,I_u)]}{\Cost(\optimalAlgo,I_u)}=\Omega(\rho),
$$
which proves the claimed competitive-ratio lower bound.
\end{proof}

\section{Estimation of \texorpdfstring{\boldmath{$n$}}{n}}\label{sec:estimate-n}

We treat population estimation as a modular subroutine: the analysis of \mainAlgorithmAcrynom relies only on the properties established in Lemmas~\ref{lem:init} and~\ref{lem:estimate}, rather than on a particular estimation procedure. This section establishes these properties for our estimator. Although geometric techniques for estimating population size are well established \cite{jurdzinski2002energy}, our analysis must also account for packet departures and the effect of rare estimation failures on expected costs.

\subsection{Population Estimation in \mainAlgorithmAcrynom}\label{sec:estimate-interface}

Whenever \mainAlgorithmAcrynom invokes \estimateFunction, all active packets participate simultaneously on the slow channel, using independent random choices. Packets that successfully transmit during estimation complete their transmissions and leave the system. The packets remaining when estimation finishes receive the same estimate at a common slow-slot boundary.

The initial estimation requires no prior bound on the population and terminates almost surely. Subsequent estimations use the initial estimate $\widehat n$ to determine the range of population sizes to consider. Estimation proceeds on the slow channel even when the fast channel becomes available. However, if \mainAlgorithmAcrynom reaches its time cutoff during a subsequent estimation, the procedure is interrupted and cleanup begins.

\subsection{Estimation Procedure and Guarantees}\label{sec:estimate-implementation}

The estimator determines the approximate number of active packets by observing how frequently the slow channel is empty. When packets transmit with relatively high probability, most slots are nonempty. As their transmission probability decreases, empty slots become more frequent. The estimator uses this relationship to identify an approximate population size.

Specifically, the estimator tests a sequence of possible population sizes, called \defn{scales}. At scale $v$, each active packet independently transmits with probability $1/v$ in each of a prescribed number of slow-channel slots, called \defn{test slots}. A scale \defn{passes} if at least half of its test slots are empty. Since all active packets observe the same ternary feedback, those remaining agree on which scale first passes and therefore on the resulting estimate.

\medskip
\noindent{\bf Initial estimation.}
The estimator tests scales $v_j=2^j$, for $j=0,1,\ldots$. At each scale $j$, it uses:
$$
h_j=2\ceil{C(j+1)/2}
$$
test slots, where $C$ is a sufficiently large constant. The estimator accepts the first passing scale and returns $\widehat n=4v_j$, setting $\widehat n_{\mathrm{cur}}=\widehat n$.

If all packets finish during estimation, no estimate is needed. For purposes of analysis, we treat the remaining test slots at that scale as empty, so that an estimate is defined even if all packets have finished.

Recall that $G_0$ is the event that the initial population estimate satisfies $n\leq\widehat n\leq32n$ and is obtained in $O(\log^2 n)$ slow slots.

\begin{lemma}\label{lem:init}
The initial call to \estimateFunction\ terminates almost surely.\footnote{This means termination occurs with probability one, even though there is no deterministic upper bound on its running time.} Let $\tau_0$ be its duration measured in slow slots. For any fixed $D>0$, a sufficiently large choice of $C$ gives an event $G_0$ with probability at least $1-O(n^{-D})$ on which the call returns an estimate $n\leq\widehat n\leq32n$ in $O(\log^2 n)$ slow slots. Since at most one packet succeeds per slow slot, this running-time bound implies that at least $n/2$ packets remain active on $G_0$, for sufficiently large $n$. The estimate is therefore also within a factor of $64$ of the number of packets remaining.

Executions outside $G_0$ contribute negligibly to the expected estimate and the expected running time; specifically,  $E[\widehat n\mid\neg G_0]\Pr(\neg G_0)=O(n^{-D})$ and 
$E[\tau_0\mid\neg G_0]\Pr(\neg G_0)=O(n^{-D})$. 
\end{lemma}

\begin{proof}
We first bound the probability that the estimator accepts a scale that is too small. The total number of test slots at scales $v\leq8n$ is $O(\log^2 n)=o(n)$. Since at most one packet succeeds per test slot, at least $n/2$ packets remain active throughout these scales, for sufficiently large $n$.

At a scale $v=2^j<n/4$, the probability that a test slot is empty is at most $e^{-n/(2v)}$, regardless of the preceding test outcomes. For the scale to pass, at least $h_j/2$ test slots must be empty. Taking a union bound over the possible choices of these slots gives:
$$
\Pr(\text{scale }j\text{ passes}\mid\text{scale }j\text{ is reached})
\leq 2^{h_j}e^{-h_jn/(4v)}
\leq e^{-h_jn/(16v)}.
$$
The last inequality holds because $n/v\geq4$. Since $h_j\geq C(j+1)$ and $(j+1)n/2^j=\Omega(\log n)$ throughout this range, choosing $C$ sufficiently large makes the probability that any such scale passes smaller than any prescribed inverse polynomial in $n$.

We next bound the probability that the estimator continues testing scales that are too large. Let $j_*$ be the first index satisfying $2^{j_*}\geq4n$. At any scale $j\geq j_*$, at most $n$ packets remain active, so the probability that a test slot is nonempty is at most $n/2^j\leq 1/4$.

This bound holds regardless of the preceding test outcomes. Thus, the number of nonempty slots satisfies the same Chernoff upper bound as $h_j$ independent tests, each nonempty with probability $1/4$. A scale fails only if more than half its test slots are nonempty, so its failure probability is $e^{-\Omega(h_j)}$. Applying these bounds successively, without assuming independence between scales, gives:
$$
\Pr(\text{reach scale }j_*+t)
\leq\exp\bigl(-\Omega(C(t\log n+t^2))\bigr)
$$
for $t\geq1$. The implicit constant in the exponent is independent of $C$. Since this bound tends to zero as $t$ increases, the initial estimation procedure terminates almost surely.

Let $G_0$ be the event that the first passing scale is at least $n/4$ and has index at most $j_*$. The preceding bounds give $\Pr(\neg G_0)=O(n^{-D})$ for sufficiently large $C$. On $G_0$, the returned estimate satisfies $n\leq\widehat n\leq32n$, and estimation takes $O(\log^2 n)$ slow slots. At most this many packets can succeed during estimation. Hence, for sufficiently large $n$, at least $n/2$ packets remain active, and $\widehat n$ is between the remaining population and $64$ times that population.

It remains to bound the contributions from executions outside $G_0$. If the first passing scale satisfies $v<n/4$, then $\widehat n<n$ and $\tau_0=O(\log^2 n)$. The probability of this event can be made sufficiently small that its contribution to both the expected estimate and expected running time is $O(n^{-D})$.

The remaining possibility is that the first passing scale has index $j_*+t$ for some $t\geq1$. The corresponding estimate and running time satisfy:
$$
\widehat n=O(n2^t),
\qquad
\tau_0=O((\log n+t)^2).
$$
The probability that this scale is the first to pass is at most the probability of reaching it. Therefore, the contributions from these executions to the expected estimate and running time are bounded, respectively, by:
$$
\begin{aligned}
\sum_{t\geq1}O(n2^t)\exp\bigl(-\Omega(C(t\log n+t^2))\bigr)
&=O(n^{-D}),\\
\sum_{t\geq1}O((\log n+t)^2)\exp\bigl(-\Omega(C(t\log n+t^2))\bigr)
&=O(n^{-D}).
\end{aligned}
$$
These bounds hold for sufficiently large $C$ because the probability of reaching increasingly large scales decreases rapidly enough to outweigh the growth in the estimate and running time. Together with the bounds for scales smaller than $n/4$, this completes the proof.
\end{proof}

\medskip
\noindent{\bf Subsequent Estimations.}
After the initial estimation, the estimator uses the common estimate $\widehat n$ to determine the range of scales to test:
$$
v=1,2,4,\ldots,2^{\ceil{\log_2(4\widehat n)}}.
$$
At each scale, it uses:
$$
h=2\ceil{C_D\log(\widehat n)/2}
$$
test slots, where $C_D$ is a sufficiently large constant.

The estimator accepts the first passing scale. If no scale passes, it selects the last scale tested. In either case, the returned estimate $\widehat n_{\mathrm{cur}}$ is the smaller of the selected scale and $\widehat n$. Thus, every completed estimation returns a value between $1$ and $\widehat n$, regardless of whether the initial estimate was accurate.

\begin{lemma}\label{lem:estimate}
For $\widehat n\geq4$, each subsequent estimation procedure uses at most $\tau_{\mathrm{est}}(\widehat n)=O(\log^2\widehat n)$ slow slots. If it finishes before the time cutoff and packets remain active, they all receive the same integer estimate between $1$ and $\widehat n$. These guarantees hold even if the initial estimate is inaccurate.

Suppose $\widehat n\geq n$ and at least $2\tau_{\mathrm{est}}(\widehat n)$ packets are active when estimation begins. For any fixed $D>0$, a sufficiently large choice of $C_D$ ensures that, except with probability $\widehat n^{-D}$, any estimate returned before the time cutoff satisfies $s\leq\widehat n_{\mathrm{cur}}\leq16s$, where $s$ is the number of packets remaining active when estimation finishes.
\end{lemma}

\begin{proof}
The estimator tests $O(\log\widehat n)$ scales, each using $O(\log\widehat n)$ test slots, giving the stated running-time bound. All packets that remain active observe the same test outcomes and therefore return the same estimate. The rule limiting that estimate to $\widehat n$ guarantees that it lies between $1$ and $\widehat n$, even if the initial estimate is inaccurate.

We now assume $\widehat n\geq n$ and establish the accuracy guarantee. It suffices to analyze the estimation procedure without interruption: any estimate returned before the time cutoff is unchanged by this assumption. Let $r$ and $s$ be the numbers of active packets when estimation begins and ends, respectively. We first show that, except with probability $\widehat n^{-D}$, the estimate satisfies $s\leq\widehat n_{\mathrm{cur}}\leq8r$.

Consider a tested scale $v$. While at least $v$ packets remain active, the probability that the next test slot is nonempty is at least $1-1/e>5/8$, regardless of earlier outcomes. For this probability calculation only, once fewer than $v$ packets remain, count every remaining test slot at that scale as nonempty. The resulting sequence satisfies the same Chernoff lower-tail bound as $h$ independent tests, each nonempty with probability $5/8$. Thus, the probability that at most half of these slots are nonempty is $e^{-\Omega(h)}$.

If $s>v$, the number of active packets never falls below $v$, so this counting rule does not change any outcome. Consequently, the probability that scale $v$ passes even though more than $v$ packets remain at the end is at most $e^{-\Omega(h)}$.

Next, consider the first power of two $v\geq4r$. It satisfies $v<8r$ and is among the tested scales because $\widehat n\geq n\geq r$. At most $r$ packets are active during its test slots, so each slot is nonempty with probability at most $r/v\leq1/4$, regardless of earlier outcomes. A standard Chernoff upper-tail bound shows that, if this scale is reached, it fails to pass with probability at most $e^{-\Omega(h)}$.

There are $O(\log\widehat n)$ scales. A union bound over scales that pass despite being smaller than $s$, together with the possible failure at the first scale at least $4r$, gives total failure probability at most:
$$
O(\log\widehat n)e^{-\Omega(h)}\leq\widehat n^{-D},
$$
for sufficiently large $C_D$, since $h\geq C_D\log\widehat n$. Outside these failures, the first passing scale lies between $s$ and $8r$. Limiting the returned value to $\widehat n$ preserves these bounds because $\widehat n\geq n\geq s$.

Finally, at most one packet succeeds per test slot. Under the lemma's assumption that $r\geq2\tau_{\mathrm{est}}(\widehat n)$, we therefore have:
$$
s\geq r-\tau_{\mathrm{est}}(\widehat n)\geq r/2.
$$
Combining this with the preceding accuracy bound gives
$s\leq\widehat n_{\mathrm{cur}}\leq8r\leq16s$,
as required.
\end{proof}

\end{document}